# Optical momentum and angular momentum in complex media: From the Abraham–Minkowski debate to unusual properties of surface plasmon-polaritons

Konstantin Y. Bliokh[1,2*], Aleksandr Y. Bekshaev[1,3*], and Franco Nori[1,4]

[1]*Center for Emergent Matter Science, RIKEN, Wako-shi, Saitama 351-0198, Japan*
[2]*Nonlinear Physics Centre, RSPE, The Australian National University, Canberra, Australia*
[3]*I. I. Mechnikov National University, Dvorianska 2, Odessa, 65082, Ukraine*
[4]*Physics Department, University of Michigan, Ann Arbor, Michigan 48109-1040, USA*
[*]*These authors contributed equally to this work*

We examine the momentum and angular-momentum (AM) properties of monochromatic optical fields in dispersive and inhomogeneous isotropic media, using the Abraham- and Minkowski-type approaches, as well as the kinetic (Poynting-like) and canonical (with separate spin and orbital degrees of freedom) pictures. While the kinetic Abraham-Poynting momentum describes the energy flux and the group velocity of the wave, the Minkowski-type quantities, with proper dispersion corrections, describe the actual momentum and angular momentum carried by the wave. The kinetic Minkowski-type momentum and AM densities agree with phenomenological results derived by Philbin. Using the canonical spin-orbital decomposition, previously used for free-space fields, we find the corresponding *canonical* momentum, spin and orbital AM of light in a dispersive inhomogeneous medium. These acquire a very natural form analogous to the Brillouin energy density and are valid for arbitrary structured fields. The general theory is applied to a non-trivial example of a surface plasmon-polariton (SPP) wave at a metal-vacuum interface. We show that the integral momentum of the SPP per particle corresponds to the SPP wave vector, and hence exceeds the momentum of a photon in the vacuum. We also provide the first accurate calculation of the transverse spin and orbital AM of the SPP. While the intrinsic orbital AM vanishes, the transverse spin can change its sign depending on the SPP frequency. Importantly, we present both macroscopic and *microscopic* calculations, thereby proving the validity of the general phenomenological results. The microscopic theory also predicts a transverse magnetization in the metal (i.e., a magnetic moment for the SPP) as well as the corresponding direct magnetization current, which provides the difference between the Abraham and Minkowski momenta.

## 1. Introduction and overview

*1.1. Abraham and Minkowski momenta*

The characterization of the momentum and angular momentum (AM) of light in continuous media is a long-standing problem, with the Abraham-Minkowski discussion in its center; see [1–5] for reviews. Although recently there were several works claiming the "resolution" of the Abraham-Minkowski controversy [5–7], debates on various aspects of optical momentum in media are still continuing. Naturally, different momenta can manifest in different types of problems or experiments. For example, one can investigate optical *forces* acting on the medium or on a small material probe in the medium [7–10]. On the other hand, one may wonder about the momentum carried by the wave *per se* and characterizing *wave* parameters such as the velocity of its propagation (phase or group) or wave vector [11–19]. In this work we mostly stick with the second approach. Throughout this paper we consider monochromatic waves with fixed frequency $\omega$, lossless isotropic media described by permittivity and permeability $\varepsilon$ and $\mu$ (which can depend on $\omega$ in the dispersive case), and cycle-averaged dynamical properties (energy, momentum, and AM) of waves.

To recall the basics of the problem, we start with the Poynting momentum density of monochromatic light in free space [20–22]:

$$\mathcal{P}_0 = g\, k_0 \operatorname{Re}\left(\mathbf{E}^* \times \mathbf{H}\right). \tag{1.1}$$

Hereafter, we use Gaussian units[1] with $g = (8\pi\omega)^{-1}$, $k_0 = \omega/c$, and all free-space quantities are marked by the subscript "0". In a non-dispersive medium, the Abraham and Minkowski momentum densities are given by [1–5]

$$\mathcal{P}_A = \mathcal{P}_0, \quad \mathcal{P}_M = \varepsilon\mu\mathcal{P}_0. \tag{1.2}$$

These two momenta are often interpreted as "*kinetic*" and "*canonical*" momenta of light, respectively [3–5,7,11,13,18]. In dispersive media, the Abraham momentum preserves its form, while the "canonical" momentum should be modified with dispersion-related terms [12–14,16,17]:

$$\tilde{\mathcal{P}}_M = \mathcal{P}_M + \{\text{dispers. terms}\}, \tag{1.3}$$

and the "naïve" Minkowski momentum (1.2) does not make physical sense. Hereafter, we mark by tilde all quantities modified by the presence of dispersion. For simplicity, we will refer to the momentum (1.3) as to the properly-modified Minkowski momentum in a dispersive medium.

For the simplest optical fields – plane waves – in a transparent medium, the Abraham and Minkowski momenta "per photon" are reduced to the following simple form [3–5,7,11–14,16–18]:

$$\mathcal{P}_A = \frac{1}{n_p n_g} \hbar \mathbf{k} = \frac{\hbar\omega}{c^2} \mathbf{v}_g, \quad \tilde{\mathcal{P}}_M = \hbar \mathbf{k}, \tag{1.4}$$

where $n_p = \sqrt{\varepsilon\mu}$ and $n_g = n_p + \omega\, dn_p/d\omega$ are the phase and group refractive indices of the medium, respectively, $\mathbf{v}_g = \partial\omega/\partial\mathbf{k}$ is the group velocity in the medium ($v_g = c/n_g$), and $\mathbf{k}$ is the wave vector in the medium, with magnitude $k = n_p k_0$.

Equations (1.4) shed light on the physical meaning of the Abraham and Minkowski momenta, which can be associated with the *group velocity* and *wave vector* in the medium. This correspondence is very general [11,18]. For example, even for inhomogeneous waves in non-transparent inhomogeneous media, such as surface plasmon-polaritons (SPPs) at metal-vacuum interfaces [23], the group velocity is still determined by the integral value of the Poynting (=Abraham) momentum [24]. At the same time, the conservation of the wave momentum and momentum matching in various resonant problems involve Minkowski momentum, i.e., the wave vector in the medium. The most known example is Snell's law in the light refraction at planar interfaces [20,22]. Furthermore, more subtle spin and orbital Hall effects (transverse beam shifts) in the refraction at planar interfaces are intimately related to the conservation of the corresponding Minkowski AM based on the same wavevector $\mathbf{k}$ [25–30].

---

[1] For the conversion between Gaussian and SI units see Appendix 4 in [20]. For the quantities used in this work, this conversion is realized via $\mathbf{E} \to \sqrt{4\pi\varepsilon_0}\,\mathbf{E}$, $\mathbf{H} \to \sqrt{4\pi\mu_0}\,\mathbf{H}$, $c \to 1/\sqrt{\varepsilon_0\mu_0}$, $\mathbf{M} \to \sqrt{\mu_0/4\pi}\,\mathbf{M}$, and $\{e, \mathbf{j}, \mathbf{d}\} \to \{e, \mathbf{j}, \mathbf{d}\}/\sqrt{4\pi\varepsilon_0}$.

However, there are problems, where the wave vector and corresponding momentum conservation are well defined and observable, while using the Minkowski momentum faces difficulties. For instance, in evanescent or surface waves, such as SPPs, the wave vector **k** exceeds $k_0$ in absolute value, and this "*super-momentum*" higher than $\hbar k_0$ per photon is observable in momentum-transfer experiments [23,31–35]. As we show below, the modified Minkowski momentum (1.3) can explain these features when integrated in localized SPP waves, but not locally in evanescent and other structured fields. This problem is related to another optical momentum dilemma.

*1.2. Canonical and kinetic pictures in free-space*

Besides the Abraham-Minkowski debate, the momentum and AM of light allow various descriptions even in *free space*. There, it is also related to the "*kinetic*" and "*canonical*" quantities, but in a different sense. Namely, the well-known Poynting momentum (1.1) corresponds to the *kinetic* momentum density, which appears in the *symmetrized (kinetic) energy-momentum tensor* of the electromagnetic field [36]. In this approach, the total AM density is determined by the same Poynting vector [20–22,36]:

$$\mathcal{J}_0 = \mathbf{r} \times \mathcal{P}_0. \tag{1.5}$$

Despite its universal character, this formalism has several practical drawbacks. First, the Poynting momentum and AM do not describe separately *spin and orbital degrees of freedom* of light. In particular, the AM density (1.5) is extrinsic, i.e., depending on the choice of the coordinate origin, and characterizing the spin (intrinsic) AM density is problematic in the Poynting formalism. At the same time, spin and orbital AM are widely explored as independent degrees of freedom in modern optics [37–43] and also in quantum field theory [44,45]. Second, the Poynting vector looses its clear physical meaning in the case of *structured* (i.e., inhomogeneous) optical fields, where it cannot explain local momentum transfer (including "super-momentum") and optical radiation-pressure forces [31–35,41,46,47].

The spin-orbital decomposition of the AM of light, described in [48–54], appears naturally in the *canonical* approach [35,36,41,44,45,47,54,55]. Using the formalism dual-symmetric with respect to the electric and magnetic fields [35,41,52,54,55–57], the *canonical momentum density* of a monochromatic light in free space is given by [35,41,54–56]

$$\mathbf{P}_0 = \frac{g}{2} \operatorname{Im} \left[ \mathbf{E}^* \cdot (\nabla) \mathbf{E} + \mathbf{H}^* \cdot (\nabla) \mathbf{H} \right]. \tag{1.6}$$

This momentum describes only the orbital part of the AM, $\mathbf{L}_0$, while the spin part is provided by an independent intrinsic quantity $\mathbf{S}_0$ [35,41,52,54,55,57]:

$$\mathbf{L}_0 = \mathbf{r} \times \mathbf{P}_0, \quad \mathbf{S}_0 = \frac{g}{2} \operatorname{Im} \left( \mathbf{E}^* \times \mathbf{E} + \mathbf{H}^* \times \mathbf{H} \right). \tag{1.7}$$

Recently, it was shown that the canonical quantities (1.6) and (1.7) are much more suitable for the description of the momentum and AM properties of free-space light than the kinetic Poynting characteristics (1.1) and (1.5). In particular, the optical force and torque on a small electric-dipole particle or an atom are given by the electric parts of the canonical momentum (1.6) and spin AM (1.7), respectively [35,41,46,47,58–60]. This makes canonical quantities directly measurable and immediately explaining numerous experiments involving spin/orbital AM [35,41,61–63] and structured light fields [31–35,46,47,64]. Moreover, using canonical formalism enabled prediction and description of unusual phenomena, such as unusual *transverse spin AM* in evanescent and other structured fields [35,41–43,46,65–68] and *super-momentum* transfer higher than $\hbar k_0$ per photon [31–35].

The canonical momentum density (1.6) can be written as a local expectation value of the quantum-mechanical momentum operator $\hat{\mathbf{p}} = -i\nabla$, and hence can be associated with the local phase gradient or *local wave vector* $\mathbf{k}_{loc}$ of the field [33,56]. This elucidates its canonical character, akin to Minkowski momentum (1.4). However, in contrast to Eq. (1.4), valid for a single plane wave, canonical momentum (1.6) describes the local phase gradient in *an arbitrary structured field*, which can consist of multiple plane waves propagating in different directions. In turn, the canonical spin AM density (1.7) describes the *local ellipticity* of the 3D polarization of an arbitrary structured field.

The relation between the kinetic (Poynting) momentum $\mathcal{P}_0$ and canonical momentum $\mathbf{P}_0$ in free space is given by the *spin-orbital momentum decomposition* [35,41,47,55,56]:

$$\mathcal{P}_0 = \mathbf{P}_0 + \mathbf{P}_0^S, \quad \mathbf{P}_0^S = \frac{1}{2}\nabla \times \mathbf{S}_0. \tag{1.8}$$

Here the canonical momentum $\mathbf{P}_0$ describes the *orbital* part (which determines the orbital AM $\mathbf{L}_0$), while the *spin momentum* $\mathbf{P}_0^S$ is related to the spin AM $\mathbf{S}_0$ via the nonlocal relation $\frac{1}{2}\int \mathbf{r} \times (\nabla \times \mathbf{S}_0) dV = \int \mathbf{S}_0 \, dV$, valid for any localized fields vanishing at infinity. Importantly, the spin momentum vanishes for plane waves and does not contribute to the *integral* (expectation) value of the wave momentum for localized fields, so that the integral kinetic and canonical momenta coincide:

$$\langle \mathcal{P}_0 \rangle = \langle \mathbf{P}_0 \rangle. \tag{1.9}$$

Here, $\langle ... \rangle = \int ... dV$, and hereafter it denotes suitable spatial integrals for localized fields.

In terms of relativistic field theory, the canonical momentum and spin densities (1.6) and (1.7) originate from the *canonical energy-momentum and AM tensors*, which are directly obtained from Noether's theorem applied to the electromagnetic field Lagrangian [36,44,45,54,55]. Two points should be emphasized here. First, the original form of these canonical tensors involves the *gauge-dependent* electromagnetic vector potential $\mathbf{A}$. The standard procedure in this case is to consider only the "transverse" (i.e., divergence-free) gauge-invariant part of this potential, $\mathbf{A}_\perp$, which for monochromatic fields is expressed via the wave electric field $\mathbf{A}_\perp = -i(c/\omega)\mathbf{E}$ [44,45,47,48–50,52,54,55,60]. Second, the standard electromagnetic-field Lagrangian is *not dual-symmetric* with respect to the electric and magnetic fields. Due to this, it results in double electric-field parts of quantities (1.6) and (1.7), with no magnetic-field parts [36,44,47,60]. However, an alternative Lagrangian formalism, dual-symmetrized between electric and magnetic contributions [55,69,70], produces the symmetric quantities (1.6) and (1.7), more natural for free Maxwell fields [35,41,54–57,59]. In this paper we employ the dual-symmetric formalism [55] and show that it is more consistent with the canonical optical properties in media than the dual-asymmetric (electric-biased) approach.

While the symmetrized (kinetic) energy-momentum tensor contains only the Poynting momentum density $\mathcal{P}_0$, the canonical energy-momentum tensor is non-symmetric, and contains both the Poynting vector $\mathcal{P}_0$, acting as the *energy flux density*, and the canonical momentum density $\mathbf{P}_0$. Remarkably, considering the energy-momentum tensor for electromagnetic waves in a medium, Dewar [11] and later Dodin and Fisch [18] found that the electromagnetic energy-momentum tensor in a (non-dispersive) medium can be modified to the Minkowski form, where the Poynting energy flux and canonical momentum are substituted by the Abraham (Poynting) and Minkowski momenta, $\mathcal{P}_A$ and $\mathcal{P}_M$, respectively. Schematically, these different forms of the energy-momentum tensors (EMTs) in free space and in media can be presented as follows:

$$\underbrace{\begin{pmatrix} \text{energy} & \text{energy flux} \\ \text{momentum} & \text{stress tensor} \end{pmatrix}}_{\text{Energy-momentum tensor (EMT)}} \rightarrow \underbrace{\begin{pmatrix} W_0 & c\mathcal{P}_0 \\ c\mathbf{P}_0 & ... \end{pmatrix}}_{\substack{\text{Canonical EMT} \\ \text{in free space} \\ \text{e.g. Soper (1976) [36]} \\ \text{Bliokh et al. (2013) [55]}}} \rightarrow \underbrace{\begin{pmatrix} \omega & \frac{\omega}{c}\mathbf{v}_g \\ c\mathbf{k} & \mathbf{k}\mathbf{v}_g \end{pmatrix}}_{\substack{\text{Plane-wave-like form} \\ \text{Dewar (1977) [11]} \\ \text{Dodin \& Fisch (2012) [18]}}} \leftarrow \underbrace{\begin{pmatrix} W & c\mathcal{P}_A \\ c\mathcal{P}_M & ... \end{pmatrix}}_{\substack{\text{Modified canonical EMT} \\ \text{in a dispersionless medium} \\ \text{Dewar (1977) [11]}}}, \quad (1.10)$$

This provides a qualitative link between the Abraham-Minkowski and kinetic-canonical (in the relativistic field-theory sense) dilemmas.

Summarizing the above considerations, one should associate the Poynting-Abraham quantities with the *energy flux* and *group velocity* of the wave-packet propagation, while the canonical and Minkowski quantities are related to the *momentum density* carried by the wave and its *wave-vector* characteristics. At the same time, we emphasize that the kinetic-canonical dilemma between $\mathcal{P}_0$ and $\mathbf{P}_0$ in the sense of relativistic field theory originates from the separation of the *spin and orbital* degrees of freedom, while the Abraham-Minkowski dilemma between $\mathcal{P}_A$ and $\mathcal{P}_M$ is related to the separation of the *medium and field* contributions to the momentum. Therefore, one can consider the spin-orbital separation in both Abraham and Minkowski momenta in a medium, as well as Abraham and Minkowski forms of the kinetic and canonical (orbital) momenta in a medium, i.e., *four* types of momenta in the medium. In what follows, we use the "kinetic" and "canonical" characteristics in the field-theory sense of the spin-orbital separation, also explicitly indicating the Abraham- and Minkowski-type quantities.

*1.3. About this work*

Here we aim to provide a complete Abraham-Minkowski and kinetic-canonical picture of optical momentum and AM in dispersive and inhomogeneous (but isotropic and lossless) media. For the reader's convenience, we summarize all the quantities under discussion in Table I, indicating their forms in free space, dispersion-free, and dispersive media. Our main emphasis in this work is on the *Minkowski*-type quantities, because these correspond to the actual wave momentum, spin, and AM in the medium, in contrast to the Abraham-type energy flux properties.

The paper is organized as follows. In this introductory Section 1 we provided a general overview of the problem. In Section 2, we discuss the general momentum and AM expressions listed in Table I and their properties. Then, in Section 3, we consider an explicit example of a SPP wave at a metal-vacuum interface.

This example (which to the best of our knowledge has never been considered in the Abraham-Minkowski context) provides a perfect test for the momentum and AM properties of structured optical fields in dispersive inhomogeneous media. The multiple advantages of the SPP system are as follows:

(i) SPP waves are well studied and readily achievable experimentally;
(ii) Even a single SPP wave is a *structured field*, for which the simplified plane-wave Eqs. (1.4) are not applicable;
(iii) SPPs exist at interfaces, i.e., in essentially *inhomogeneous* media.
(iv) SPPs exhibit nontrivial momentum and AM properties, including *super-momentum* [31–35] and *transverse spin AM* [35,41–43,46,65–68];
(v) *Dispersion* of the metal is crucial for the SPP properties.

Thus, SPPs provide both an accessible and highly nontrivial system to study optical momentum and AM.

In Section 3 we apply the general expressions from Table I to calculate the momentum and AM properties of SPPs. In particular, we present the first accurate calculations of the canonical and Minkowski momenta, as well as the transverse spin and orbital AM of a SPP. Notably, even the *integral* canonical or Minkowski momentum of SPP exceeds $\hbar k_0$ per particle, and this offers the first example of the *integral super-momentum* (previously known only locally). Moreover, the intrinsic orbital AM of the SPP vanishes, whereas the integral transverse spin AM can change its sign depending on parameters.

| | Kinetic picture | | Canonical (spin-orbital) picture | |
|---|---|---|---|---|
| | Abraham (energy flux) | Minkowski (wave moment.) | Abraham (energy flux) | Minkowski (wave momentum) |
| $\varepsilon = 1$ $\mu = 1$ | $\mathcal{P}_0 \propto k_0 \, \text{Re}\left(\mathbf{E}^* \times \mathbf{H}\right)$ $\mathcal{J}_0 = \mathbf{r} \times \mathcal{P}_0$ | | $\mathbf{P}_0 \propto \frac{1}{2} \text{Im}\left[\mathbf{E}^* \cdot (\nabla)\mathbf{E} + \mathbf{H}^* \cdot (\nabla)\mathbf{H}\right]$ $\mathbf{S}_0 \propto \frac{1}{2} \text{Im}\left(\mathbf{E}^* \times \mathbf{E} + \mathbf{H}^* \times \mathbf{H}\right)$ *Berry (2009)* [56], *Barnett (2010)* [52], *Bliokh et al. (2013)* [55] | |
| $\varepsilon$ $\mu$ | $\mathcal{P}_A = \mathcal{P}_0$ $\mathcal{J}_A = \mathcal{J}_0$ | $\mathcal{P}_M = \varepsilon \mu \mathcal{P}_0$ $\mathcal{J}_M = \mathbf{r} \times \mathcal{P}_M = \varepsilon \mu \mathcal{J}_0$ | $\mathbf{P}_A \propto$ $\text{Im}\left[\frac{\mathbf{E}^* \cdot (\nabla)\mathbf{E}}{2\mu} + \frac{\mathbf{H}^* \cdot (\nabla)\mathbf{H}}{2\varepsilon}\right]$ $+\{\text{grad.}\}$ $\mathbf{S}_A \propto \text{Im}\left(\frac{\mathbf{E}^* \times \mathbf{E}}{2\mu} + \frac{\mathbf{H}^* \times \mathbf{H}}{2\varepsilon}\right)$ *Bliokh et al. (2012,2014)* [65,35] | $\mathbf{P}_M \propto$ $\text{Im}\left[\frac{\varepsilon \mathbf{E}^* \cdot (\nabla)\mathbf{E}}{2} + \frac{\mu \mathbf{H}^* \cdot (\nabla)\mathbf{H}}{2}\right]$ $\mathbf{S}_M \propto \text{Im}\left(\frac{\varepsilon \mathbf{E}^* \times \mathbf{E}}{2} + \frac{\mu \mathbf{H}^* \times \mathbf{H}}{2}\right)$ |
| $\varepsilon(\omega)$ $\mu(\omega)$ | | $\tilde{\mathcal{P}}_M = \mathcal{P}_M + \{\text{dispers.}\}$ $\tilde{\mathcal{J}}_M = \mathbf{r} \times \tilde{\mathcal{P}}_M + \{\text{dispers.}\}$ *Philbin (2012)* [16,17], *this work* | | $\tilde{\mathbf{P}}_M \propto$ $\text{Im}\left[\frac{\tilde{\varepsilon} \mathbf{E}^* \cdot (\nabla)\mathbf{E}}{2} + \frac{\tilde{\mu} \mathbf{H}^* \cdot (\nabla)\mathbf{H}}{2}\right]$ $\tilde{\mathbf{S}}_M \propto \text{Im}\left(\frac{\tilde{\varepsilon} \mathbf{E}^* \times \mathbf{E}}{2} + \frac{\tilde{\mu} \mathbf{H}^* \times \mathbf{H}}{2}\right)$ *this work* |

**Table I.** Four possible pictures of the optical momentum and angular momentum (AM) densities in free space, non-dispersive isotropic media, and dispersive (generally inhomogeneous) media. The Abraham- and Minkowski-type, kinetic and canonical (spin-orbital) quantities are shown. In all cases, the *kinetic Abraham-Poynting* momentum density $\mathcal{P}_A = \mathcal{P}_0$ describes the energy flux and group velocity of the wave, whereas the *canonical Minkowski-type* momentum and spin densities $\tilde{\mathbf{P}}_M$ and $\tilde{\mathbf{S}}_M$ provide a clear and self-consistent picture of the momentum and angular momentum carried by the wave. In turn, the kinetic-Minkowski and canonical-Abraham quantities have less natural forms with cumbersome dispersive and gradient corrections (indicated as {dispers.} and {grad.} here).

Importantly, in Section 4, we provide *microscopic* calculations of the momentum and AM in SPPs. Taking into account both microscopic electromagnetic fields and the motion of free electrons in the metal, we obtain the Minkowski and canonical quantities previously introduced using macroscopic phenomenological considerations. This validates the use of these quantities for structured optical fields in dispersive and inhomogeneous media. Moreover, the microscopic theory predicts a *transverse magnetization* in the metal (i.e., a magnetic moment for the SPP) as well as the corresponding *direct magnetization current*, which corresponds to the difference between the Abraham and Minkowski-type momenta.

Finally, in Section 5, we briefly discuss issues related to the dual symmetry between electric- and magnetic-field contributions. We show that while integral electric and magnetic conributions to the momentum and spin are equal for localized fields in free space [52,55], this is not the case for localized fields in media. Most importantly, we find that the microscopic calculations of Section 4 are only compatible with the *dual-symmetric* (rather than electric-biased) forms of the canonical quantities.

We should also briefly mention preceding works, which considered some of the above aspects of SPPs. First, Nakamura [71] performed *microscopic* calculations of the transverse AM of SPPs. Although results of that work are erroneous in several aspects (calculation errors, mixing of the spin and orbital AM, etc.), its methodology inspired us to perform microscopic calculations presented in Section 4. Second, Kim and Wang aimed to calculated Abraham and "naïve" Minkowski (without dispersion terms) versions of the transverse spin AM in SPPs [72–74]. However, their results are also misleading. First, the definitions in [72–74] do not describe the Abraham-type spin, which was properly defined and calculated in [35,65], and which corresponds to an energy-flux property rather than the actual spin AM carried by the wave. Second, the "naïve" Minkowski expressions are not applicable to waves in dispersive media and lead to erroneous results. Thus, the first accurate calculation of the transverse spin and other "canonical" characteristics of SPPs are provided in the present work.

## 2. Momentum and angular momentum of light in dispersive inhomogeneous media

Throughout this paper we consider monochromatic electric and magnetic fields: $\mathcal{E}(\mathbf{r},t) = \text{Re}\left[\mathbf{E}(\mathbf{r})e^{-i\omega t}\right]$ and $\mathcal{H}(\mathbf{r},t) = \text{Re}\left[\mathbf{H}(\mathbf{r})e^{-i\omega t}\right]$. The main independent dynamical properties of light are: energy, momentum, as well as spin and orbital angular momentum (AM). One can also add here helicity, which is an independent conserved quantity [55,57,59,69,70,75–79]. The momentum and AM characteristics, both kinetic and canonical, for monochromatic free-space fields are given by Eqs. (1.1) and (1.5)–(1.8). For completeness, here we add the energy density [20–22]

$$W_0 = \frac{g\omega}{2}\left(|\mathbf{E}|^2 + |\mathbf{H}|^2\right). \tag{2.1}$$

We now consider an isotropic lossless dispersive and inhomogeneous medium, which is characterized by the real frequency-dependent permittivity $\varepsilon(\omega,\mathbf{r})$ and permeability $\mu(\omega,\mathbf{r})$. In this case, the complex field amplitudes satisfy stationary Maxwell equations

$$\nabla \cdot (\mu \mathbf{H}) = 0, \quad \mu \mathbf{H} = -\frac{i}{k_0}\nabla \times \mathbf{E},$$

$$\nabla \cdot (\varepsilon \mathbf{E}) = 0, \quad \varepsilon \mathbf{E} = \frac{i}{k_0}\nabla \times \mathbf{H}. \tag{2.2}$$

Note that these source-free equations are used in the decomposition of the Poynting momentum density into canonical and spin parts, Eq. (1.8).

The energy density of a monochromatic optical field in such a medium is described by the well-known Brillouin expression [20,22]:

$$\boxed{\tilde{W} = \frac{g\omega}{2}\left(\tilde{\varepsilon}|\mathbf{E}|^2 + \tilde{\mu}|\mathbf{H}|^2\right)}, \tag{2.3}$$

where

$$\tilde{\varepsilon} = \varepsilon + \omega\frac{d\varepsilon}{d\omega}, \quad \tilde{\mu} = \mu + \omega\frac{d\mu}{d\omega}. \tag{2.4}$$

Describing the optical momentum density in a medium is a more sophisticated problem. On the one hand, the Abraham momentum $\boldsymbol{\mathcal{P}}_A$ preserves its Poynting-vector form in the medium, Eq. (1.2). By analogy with the canonical decomposition (1.8), one can decompose it into orbital and spin parts, $\boldsymbol{\mathcal{P}}_A = \mathbf{P}_A + \mathbf{P}_A^S = \mathbf{P}_A + \frac{1}{2}\nabla \times \mathbf{S}_A$, where

$$\mathbf{P}_A = \frac{g}{2}\mathrm{Im}\left[\mu^{-1}\mathbf{E}^* \cdot (\nabla)\mathbf{E} + \varepsilon^{-1}\mathbf{H}^* \cdot (\nabla)\mathbf{H}\right] - \frac{g}{4}\left[\nabla \mu^{-1} \times \mathrm{Im}\left(\mathbf{E}^* \times \mathbf{E}\right) + \nabla \varepsilon^{-1} \times \mathrm{Im}\left(\mathbf{H}^* \times \mathbf{H}\right)\right], \quad (2.5)$$

$$\mathbf{S}_A = \frac{g}{2}\mathrm{Im}\left(\mu^{-1}\mathbf{E}^* \times \mathbf{E} + \varepsilon^{-1}\mathbf{H}^* \times \mathbf{H}\right), \quad (2.6)$$

This spin-orbital decomposition was introduced in [65] and was used in [35,46,80] because of its convenience in homogeneous media. However, in inhomogeneous media, the canonical momentum density (2.5) acquires cumbersome gradient terms. Moreover, the physical interpretation of the quantities (2.5) and (2.6) is not quite clear. Indeed, as we discussed above, the Abraham momentum $\boldsymbol{\mathcal{P}}_A$ should be associated with the *energy flux* density and *group velocity*, Eq. (1.10), rather than with the wave momentum density. Therefore, the Abraham-type quantities (2.5) and (2.6) correspond to the orbital and spin parts of the energy flux density, but cannot be regarded as canonical momentum and spin densities in the wave. In addition, at interfaces between media, the Abraham-type spin density (2.6) is discontinuous, and the corresponding gradient terms in Eq. (2.5) produce *singular* delta-function contributions to the canonical and spin momentum densities $\mathbf{P}_A$ and $\mathbf{P}_A^S$ [65]. This makes the Abraham-type spin-orbital decomposition imperfect. Note also that, similarly to the free-space Eqs. (1.8) and (1.9), the solenoidal spin part of the energy flux does not contribute to the plane-wave and integral characteristics:

$$\langle \boldsymbol{\mathcal{P}}_A \rangle = \langle \mathbf{P}_A \rangle. \quad (2.7)$$

Therefore, in some plane-wave or integral calculations it could be more convenient to use Eq. (2.5) as the energy flux density.

To describe physically-meaningful momentum and AM densities in the optical field, one should use the Minkowski momentum. Its simple form (1.2) is not valid in the presence of dispersion, and several works discussed modifications of the Minkowski-type wave momentum in a dispersive medium [12–14,16,17]. The most general expression, suitable for structured wave fields was derived by Philbin [16,17] using the phenomenological Lagrangian formalism and Noether's theorem:

$$\boxed{\tilde{\boldsymbol{\mathcal{P}}}_M = \boldsymbol{\mathcal{P}}_M + \frac{g\omega}{2}\mathrm{Im}\left[\frac{d\varepsilon}{d\omega}\mathbf{E}^* \cdot (\nabla)\mathbf{E} + \frac{d\mu}{d\omega}\mathbf{H}^* \cdot (\nabla)\mathbf{H}\right]}. \quad (2.8)$$

Here, the first term is the "naïve" Minkowski momentum (1.2), while the second term describes the dispersion-related correction. For a plane wave in a transparent homogeneous medium, the Abraham and modified-Minkowski momenta $\boldsymbol{\mathcal{P}}_A$ and $\tilde{\boldsymbol{\mathcal{P}}}_M$ yield simplified Eqs. (1.4) [3–5,7,11–14,16–18].

Since the Minkowski-type momentum (2.8) represents "canonical" wave-vector properties of the wave, it makes sense to find the spin-orbital decomposition, similar to Eq. (1.8), and introduce the corresponding canonical spin and orbital properties. In doing so, we apply the standard Poynting-vector decomposition (1.8) to the first (Minkowski) term in Eq. (2.8) and add the second dispersive term to the orbital part (because of its natural orbital form). This results in $\tilde{\boldsymbol{\mathcal{P}}}_M = \tilde{\mathbf{P}}_M + \mathbf{P}_M^S$:

$$\tilde{\mathbf{P}}_M = \frac{g}{2}\operatorname{Im}\left[\tilde{\varepsilon}\mathbf{E}^*\cdot(\nabla)\mathbf{E} + \tilde{\mu}\mathbf{H}^*\cdot(\nabla)\mathbf{H}\right], \tag{2.9}$$

$$\mathbf{P}_M^S = \frac{1}{2}\nabla\times\mathbf{S}_M, \quad \mathbf{S}_M = \frac{g}{2}\operatorname{Im}\left(\varepsilon\mathbf{E}^*\times\mathbf{E} + \mu\mathbf{H}^*\times\mathbf{H}\right). \tag{2.10}$$

Notably, the *canonical (orbital) momentum density (2.9)* has a nice form similar to the free-space momentum (1.6) with the $\tilde{\varepsilon}$ and $\tilde{\mu}$ multipliers, exactly as in the Brillouin energy density (2.3). Furthermore, the momentum (2.9) is free of cumbersome gradient terms, present in the canonical Abraham-type momentum (2.5). However, the quantity $\mathbf{S}_M$ in Eq. (2.10) is the "naïve" Minkowski spin AM density, which lacks dispersive corrections. As we show below, this is *not* the canonical spin AM density of the wave. Notably, in the SPP example considered below, the quantity $\mathbf{S}_M$ is *continuous* at the interface, and therefore the canonical and spin parts of the modified Minkowski momentum density, $\tilde{\mathbf{P}}_M$ and $\mathbf{P}_M^S$, are *free* of delta-function singularities. This makes the Minkowski-type spin-orbital decomposition more appealing than the Abraham one.

Akin to Eqs. (1.8), (1.9), and (2.7), the solenoidal spin momentum (2.10) vanishes for plane waves and does not contribute to the integral momentum values. Therefore, the integral values of the kinetic and canonical Minkowski-type momenta (2.8) and (2.9) coincide for localized fields:

$$\langle\tilde{\boldsymbol{\mathcal{P}}}_M\rangle = \langle\tilde{\mathbf{P}}_M\rangle. \tag{2.11}$$

Thus, one can use either of these momenta in calculations of the integral or plane-wave properties.

To determine the canonical spin and orbital AM in a dispersive medium, we start with the *kinetic* (total) Minkowski-type AM. Again, the Minkowski-type analogue of Eq. (1.5) in a dispersive medium was found by Philbin and Allanson [17]:

$$\tilde{\boldsymbol{\mathcal{J}}}_M = \mathbf{r}\times\tilde{\boldsymbol{\mathcal{P}}}_M + \frac{g\omega}{2}\operatorname{Im}\left[\frac{d\varepsilon}{d\omega}\mathbf{E}^*\times\mathbf{E} + \frac{d\mu}{d\omega}\mathbf{H}^*\times\mathbf{H}\right]. \tag{2.12}$$

Importantly, the AM density (2.12) breaks the simple relation (1.5) between the kinetic momentum and AM densities, $\tilde{\boldsymbol{\mathcal{J}}}_M \neq \mathbf{r}\times\tilde{\boldsymbol{\mathcal{P}}}_M$, and contains a dispersion-related correction of the spin-like local form. Substituting $\tilde{\boldsymbol{\mathcal{P}}}_M = \tilde{\mathbf{P}}_M + \mathbf{P}_M^S$ with Eqs. (2.9) and (2.10) into Eq. (2.12) and using the nonlocal relation between the spin momentum and spin AM, $\frac{1}{2}\int\mathbf{r}\times(\nabla\times\mathbf{S}_M)dV = \int\mathbf{S}_M dV$, we derive the *canonical orbital and spin AM densities* in a dispersive medium:

$$\tilde{\mathbf{S}}_M = \frac{g}{2}\operatorname{Im}\left(\tilde{\varepsilon}\mathbf{E}^*\times\mathbf{E} + \tilde{\mu}\mathbf{H}^*\times\mathbf{H}\right), \quad \tilde{\mathbf{L}}_M = \mathbf{r}\times\tilde{\mathbf{P}}_M. \tag{2.13}$$

Here the dispersion terms from Eq. (2.12) correct the "naïve" Minkowski spin density $\mathbf{S}_M$. Remarkably, Eqs. (2.13) have a very nice form, similar to the free-space Eqs. (1.7), but now with the same $\tilde{\varepsilon}$ and $\tilde{\mu}$ multipliers as in the Brillouin energy density (2.3) and canonical momentum density (2.9). The integral values of the kinetic and canonical AM (2.12) and (2.13) coincide for localized fields:

$$\langle\tilde{\boldsymbol{\mathcal{J}}}_M\rangle = \langle\tilde{\mathbf{S}}_M\rangle + \langle\tilde{\mathbf{L}}_M\rangle. \tag{2.14}$$

Equations (2.3), (2.9), and (2.13) constitute a set of canonical characteristics of a monochromatic light in a dispersive and inhomogeneous medium. Importantly, these can be written as local expectation values of quantum-mechanical energy ($\omega$), momentum ($\hat{\mathbf{p}} = -i\nabla$), and spin-1 ($\hat{\mathbf{S}}$) operators [35,54–56,59]:

$$\tilde{W} = (\psi|\omega|\psi), \quad \tilde{\mathbf{P}}_M = \text{Re}(\psi|\hat{\mathbf{p}}|\psi), \quad \tilde{\mathbf{S}}_M = (\psi|\hat{\mathbf{S}}|\psi), \quad \tilde{\mathbf{L}}_M = \text{Re}(\psi|\mathbf{r}\times\hat{\mathbf{p}}|\psi), \quad (2.15)$$

using the following wave-function:

$$\psi = \sqrt{\frac{g}{2}}\hat{M}^{1/2}\begin{pmatrix} \mathbf{E} \\ \mathbf{H} \end{pmatrix}, \quad (\psi|\psi') = \psi^\dagger\psi, \quad \hat{M} = \begin{pmatrix} \tilde{\varepsilon} & 0 \\ 0 & \tilde{\mu} \end{pmatrix}. \quad (2.16)$$

Exactly the same formalism for electromagnetic bi-linear forms (including Berry connection and other topological characteristics) in dispersive media was recently suggested in works by Silveirinha [81–83]. Thus, the above equations bring together approaches developed by (i) Philbin (kinetic Minkowski-type momentum and AM in dispersive media) [16,17], (ii) Bliokh *et al*. (canonical momentum and AM pictures in free space) [35,41,52,55,56], and (iii) Silveirinha (electromagnetic bi-linear forms in dispersive media) [81–83].

The natural form of the energy, momentum, and AM in Eqs. (2.3), (2.9), (2.13), and (2.15) suggests that the canonical form of the Minkowski-type momentum and AM densities is more suitable for the description of the optical momentum and AM than the previously-used kinetic Minkowski-type approach, Eqs. (2.8) and (2.12), and the Abraham-type quantities (2.5) and (2.6). For example, consider a polarized plane wave propagating in a homogeneous dispersive medium. All field components have the same phase factor $\exp(i\mathbf{k}\cdot\mathbf{r})$, the electric and magnetic field amplitudes are related by $|\mathbf{E}|^2/\mu = |\mathbf{H}|^2/\varepsilon$, whereas the ellipticity of the polarization can be characterized by the helicity $\sigma$, such that $\text{Im}(\mathbf{E}^*\times\mathbf{E}) = \sigma|\mathbf{E}|^2\bar{\mathbf{k}}$ (and a similar equation for the magnetic field), where $\bar{\mathbf{k}} = \mathbf{k}/k$ characterizes the direction of the wave propagation. Using these simple properties, from Eqs. (2.3), (2.5), (2.6), (2.9) and (2.13), we readily obtain the ratios of the canonical Abraham- and Minkowski-type momentum and spin densities to the energy density of the wave:

$$\frac{\mathbf{P}_A}{\tilde{W}} = \frac{1}{n_p n_g}\frac{\mathbf{k}}{\omega}, \quad \frac{\tilde{\mathbf{P}}_M}{\tilde{W}} = \frac{\mathbf{k}}{\omega}, \quad \frac{\mathbf{S}_A}{\tilde{W}} = \frac{1}{n_p n_g}\frac{\sigma}{\omega}\bar{\mathbf{k}}, \quad \frac{\tilde{\mathbf{S}}_M}{\tilde{W}} = \frac{\sigma}{\omega}\bar{\mathbf{k}}. \quad (2.17)$$

The first two of these equations exactly correspond to Eqs. (1.4), whereas the other two equations provide their counterparts for the spin AM (cf. [17,84]). In this manner, the Minkowski-type momentum $\tilde{\mathbf{P}}_M$ and spin AM $\tilde{\mathbf{S}}_M$ correspond to the values $\hbar\mathbf{k}$ and $\hbar\sigma\bar{\mathbf{k}}$ per photon, as one would expect for photons, the Abraham-type momentum $\mathbf{P}_A$ determines the group velocity (1.4), while the Abraham-type spin $\mathbf{S}_A$ does not have a clear physical meaning.

It is important to note that it is the Minkowski-type wave momentum and AM that are *conserved* in media with the corresponding translational and rotational symmetries. First, this follows in the most general form from the results of [16,17], where the kinetic quantities (2.8) and (2.12) were derived from Noether's theorem. In view of Eqs. (2.11) and (2.14), this is also true for the canonical Minkowski-type quantities (2.9) and (2.13). Second, in the plane-wave Eqs. (2.17), the Minkowski-type momentum and spin exactly correspond to the tangent-momentum and normal-AM conservation laws in the wave refraction at an interface between two media [20,22,25–30] (Snell's law and optical beam shifts).

In this Section we presented a *macroscopic* phenomenological introduction of these quantities. Below, considering SPPs at the vacuum-metal interface, we show that this macroscopic model is in exact agreement with *microscopic* calculations taking into account separate electron and field contributions. It should be also noted that in the absence of dispersion, $\tilde{\varepsilon} = \varepsilon$, $\tilde{\mu} = \mu$, and both kinetic and canonical characteristics discussed in this Section acquire simplified Minkowski forms, shown in Table I.

## 3. Macroscopic calculations for a surface plasmon-polariton

*3.1. SPP fields and parameters.*

We now consider an explicit example of a structured optical field in a dispersive and inhomogeneous medium: a surface plasmon-polariton (SPP) at the metal-vacuum interface [23]. The geometry of the problem is shown in Fig. 1(a), where the interface is the $x = 0$ plane with the vacuum in the $x > 0$ half-space (medium 1) and metal in the $x < 0$ half-space (medium 2), whereas the SPP wave propagates along the $z$-axis with the wavevector $\mathbf{k}_p = k_p \bar{\mathbf{z}}$ (hereafter, $\bar{\mathbf{x}}$, $\bar{\mathbf{y}}$, and $\bar{\mathbf{z}}$ denote the unit vectors of the corresponding axes). The permittivity and permeability of the metal are given using the standard plasma model [23]:

$$\mu = 1, \quad \varepsilon(\omega) = 1 - \frac{\omega_p^2}{\omega^2}. \tag{3.1}$$

Here, $\omega_p^2 = 4\pi n_0 e^2 / m$ is the plasma frequency, where $n_0$ is the volume density of free electrons in the metal, $e < 0$ is the electron charge, and $m$ is the electron mass. Thus, the metal is a dispersive medium with $\tilde{\varepsilon} = 1 + \omega_p^2/\omega^2 \neq \varepsilon$. SPPs are electromagnetic surface TM waves that exist at frequencies $\omega < \omega_p / \sqrt{2}$ where $\varepsilon < -1$ [23].

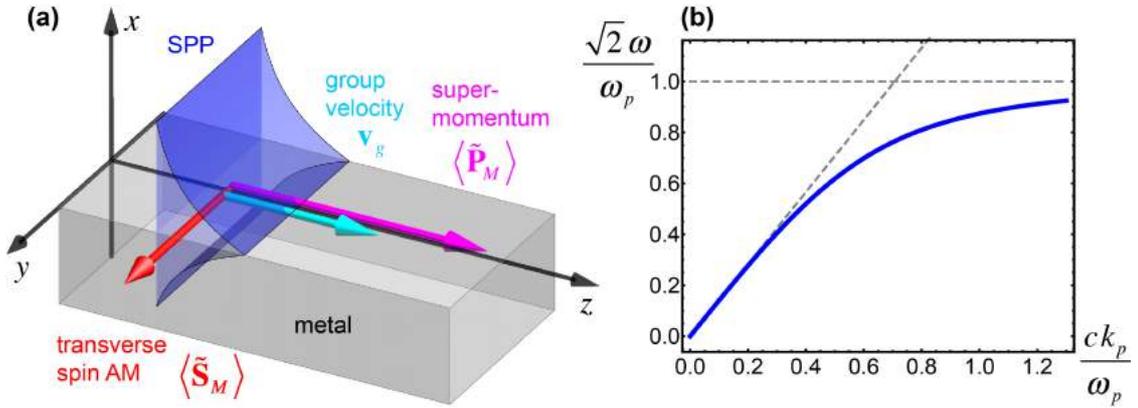

**Fig. 1**. (a) Schematic picture of a surface plasmon-polariton (SPP) wave at the metal-vacuum interface [23]. The subluminal group velocity (3.8), super-momentum (3.9), and the transverse spin (3.14) are schematically shown. (b) The dispersion of the SPP, $\omega(k_p)$, obtained from Eqs. (3.1) and (3.3).

The electric and magnetic fields of a single SPP wave can be written as [23,24,65]:

$$\mathbf{E} = A \begin{cases} \left(\bar{\mathbf{x}} - i\dfrac{\kappa_1}{k_p}\bar{\mathbf{z}}\right)\exp(ik_p z - \kappa_1 x), & x > 0 \\ \dfrac{1}{\varepsilon}\left(\bar{\mathbf{x}} + i\dfrac{\kappa_2}{k_p}\bar{\mathbf{z}}\right)\exp(ik_p z + \kappa_2 x), & x < 0 \end{cases} \tag{3.2a}$$

$$\mathbf{H} = A \begin{cases} \overline{\mathbf{y}} \dfrac{k_0}{k_p} \exp(ik_p z - \kappa_1 x), & x > 0 \\ \overline{\mathbf{y}} \dfrac{k_0}{k_p} \exp(ik_p z + \kappa_2 x), & x < 0 \end{cases} \qquad (3.2b)$$

where $A$ is the field amplitude, whereas the wave number and spatial decay constants of the SPP field are

$$k_p = \frac{\sqrt{-\varepsilon}}{\sqrt{-1-\varepsilon}} k_0, \quad \kappa_1 = \frac{1}{\sqrt{-1-\varepsilon}} k_0, \quad \kappa_2 = \frac{-\varepsilon}{\sqrt{-1-\varepsilon}} k_0. \qquad (3.3)$$

The vacuum part of the SPP fields (3.2) and (3.3) is a free-space TM-polarized evanescent wave with $k_p^2 - \kappa_1^2 = k_0^2$. From Eqs. (3.1) and (3.3), one can obtain the dependence $k_p(\omega)$ and the dispersion of the SPP, $\omega(k_p)$, shown in Fig. 1(b).

*3.2. Energy, energy flux, and momentum of SPPs.*

Substituting Eqs. (3.2) and (3.3) into Eq. (2.3), we obtain the energy density distribution in the SPP field:

$$\tilde{W} = g|A|^2 \omega \begin{cases} \exp(-2\kappa_1 x), & x > 0 \\ \dfrac{1-\varepsilon+\varepsilon^2}{\varepsilon^2} \exp(2\kappa_2 x), & x < 0 \end{cases} \qquad (3.4)$$

Note that the distribution (3.4) is discontinuous at the interface $x = 0$. Since SPP field is localized along the $x$ coordinate, we can also calculate the integral "expectation value" of the SPP energy integrating $\tilde{W}$ over $x$:

$$\left\langle \tilde{W} \right\rangle = g|A|^2 \frac{\omega}{k_p} \frac{(1-\varepsilon)(1+\varepsilon^2)}{2\varepsilon^2 \sqrt{-\varepsilon}}, \qquad (3.5)$$

where $\langle ... \rangle \equiv \int_{-\infty}^{\infty} ... dx$. In what follows, we express the integral momentum and AM characteristics of SPPs with respect to the energy (3.5), in order to highlight their values "per plasmon". From the energy-density distribution, we can also find the position of the center of energy along the $x$-axis:

$$\langle x \rangle \equiv \frac{1}{\langle \tilde{W} \rangle} \int_{-\infty}^{\infty} x \tilde{W} \, dx = -\frac{1}{2k_p} \frac{1+\varepsilon^2+\varepsilon^3}{(1+\varepsilon^2)\sqrt{-\varepsilon}}. \qquad (3.6)$$

It follows from here that for $\omega < \omega_p/1.57$ the energy centroid is located in the vacuum: $\langle x \rangle > 0$, while for $\omega_p/1.57 < \omega < \omega_p/\sqrt{2}$ it moves into the metal: $\langle x \rangle < 0$. Dependences of the integral energy (3.5) and center-of-energy position (3.6) on the SPP frequency are shown in Fig. 2.

The Abraham-Poynting momentum density, or rather the *energy flux*, Eqs. (1.1) and (1.2), for the SPP fields (3.2) and (3.3) yields

$$\mathcal{P}_A = g|A|^2 \frac{k_0^2}{k_p} \overline{\mathbf{z}} \begin{cases} \exp(-2\kappa_1 x), & x > 0 \\ \dfrac{1}{\varepsilon} \exp(2\kappa_2 x), & x < 0 \end{cases} \qquad (3.7)$$

Note that this flux is *negative* (backward) inside the metal, producing a *vortex-like* energy circulation in SPP wave packets [24,65,85]. Nonetheless, the integral energy flux $\langle \mathcal{P}_A \rangle$ is positive, and, in ratio to the energy (3.5), it determines the *group velocity* of SPPs [24,86]:

$$\boxed{\mathbf{v}_g = \frac{c^2 \langle \mathcal{P}_A \rangle}{\langle \tilde{W} \rangle} = c \frac{\sqrt{-\varepsilon}(-1-\varepsilon)^{3/2}}{1+\varepsilon^2} \bar{\mathbf{z}} = \frac{\partial \omega}{\partial k_p} \bar{\mathbf{z}}}. \qquad (3.8)$$

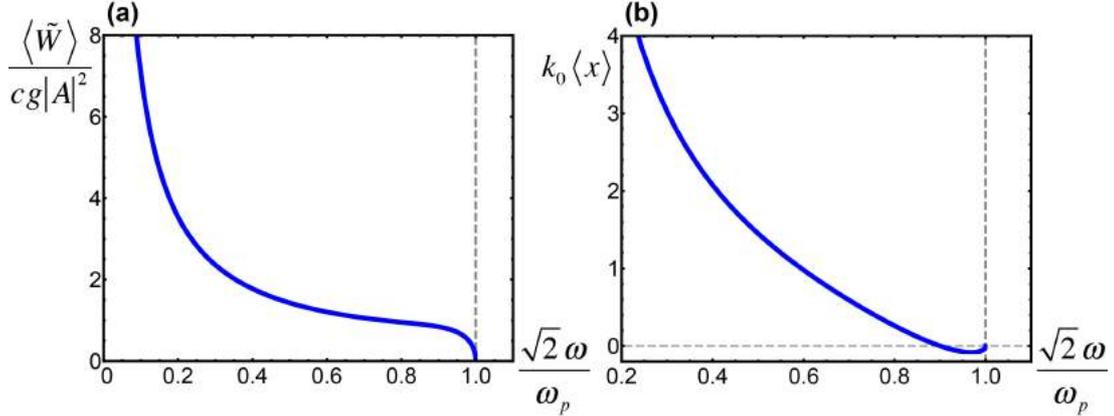

**Fig. 2**. The integral energy (3.5) and center-of-energy position (3.6) versus the SPP frequency $\omega$.

Importantly, the absolute value of the group velocity (3.8) is always *subluminal*: $v_g < c$, which corresponds to the subluminal propagation of a SPP wave packet. This also means that the integral Abraham momentum is *smaller* than $\hbar k_0$ "per plasmon". Although the plane-wave homogenous-medium equations (1.4) and (2.17) are not directly applicable to the structured SPP wave at the metal-vacuum interface, the group velocity (3.8) can be written in the form $v_g = c / n_g^{\text{eff}}$, where we introduced an effective phase refractive index for SPPs, $n_p^{\text{eff}} = k_p / k_0 > 1$, and the corresponding effective group index $n_g^{\text{eff}} = n_p^{\text{eff}} + \omega \, dn_p^{\text{eff}} / d\omega > 1$. This shows that the relations (1.4) between the Abraham energy flux, group velocity, and refractive indices are rather general and can be extended, using integral expectation values, to localized states in inhomogeneous media. The frequency dependence of the SPP group velocity (3.8) is depicted in Fig. 3(a).

The orbital and spin parts of the energy flux (3.7), $\mathbf{P}_A$ and $\mathbf{P}_A^S = \frac{1}{2} \nabla \times \mathbf{S}_A$, Eqs. (2.5) and (2.6), have been analyzed for SPPs in [65], and we do not reproduce these here. We just recall that these parts have *singular* delta-function contributions at the interface $x = 0$ due to the gradient terms in Eq. (2.5) and discontinuity of the Abraham spin (2.6) $\mathbf{S}_A$ (shown below). These singular contributions are crucial to satisfy Eq. (2.7).

We now calculate the Minkowski-type momentum for the SPP fields. These have a more natural form in the *canonical* approach. Indeed, using Eq. (2.9), we readily obtain for the field (3.2) with the common $\exp(ik_p z)$ phase factor:

$$\boxed{\tilde{\mathbf{P}}_M = k_p \frac{\tilde{W}}{\omega} \bar{\mathbf{z}}}, \quad \boxed{\langle \tilde{\mathbf{P}}_M \rangle = k_p \frac{\langle \tilde{W} \rangle}{\omega} \bar{\mathbf{z}}}. \qquad (3.9)$$

Note that the momentum density $\tilde{\mathbf{P}}_M$ is *positive* both in the vacuum and in metal, in contrast to the energy flux (3.7). Moreover, it *does not have delta-function singularities at the interface*, in contrast to the canonical Abraham-type momentum $\mathbf{P}_A$ [65].

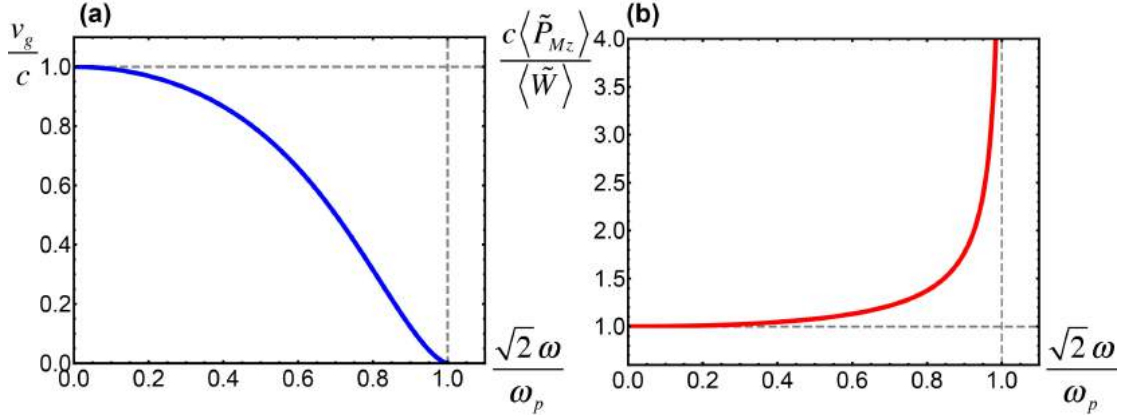

**Fig. 3**. The group velocity (3.8) and integral Minkowski-type momentum (3.9) of a SPP versus frequency. While the group velocity is always sub-luminal, $v_g < c$, the integral momentum of SPP corresponds to the *super-momentum* $\hbar k_p > \hbar k_0$ "per plasmon".

Assuming that the energy is quantized as $\hbar\omega$ "per plasmon", Eqs. (3.9) mean that *the SPP carries super-momentum $\hbar k_p > \hbar k_0$ per plasmon*. Remarkably, so far the super-momentum was described only *locally*, in evanescent waves or near optical vortices [31–35,56,87,88]. At the same time, the integral momentum for any localized wave field in free space is always less than $\hbar k_0$ per photon [33]. Equations (3.9) show an example of the *integral* super-momentum. Of course, one can say that a canonical momentum higher than $\hbar k_0$ per photon appears in any medium with phase refractive index $n_p > 1$. Equations (3.9) can also be written in the form of Eqs. (1.4) and (2.17) with $\mathbf{k} = k_p \bar{\mathbf{z}}$ and *effective* refractive index $n_p^{\text{eff}} = k_p / k_0 > 1$. However, in the case of SPP waves, this effective refractive index originates from the inhomogeneous evanescent character of surface waves rather than from the high permittivity of the medium. Indeed, in the limit $\omega \to \omega_p / \sqrt{2}$, we have $n_p^{\text{eff}} \to \infty$, while $\varepsilon \to -1$. Note that although in some works the super-momentum was interpreted as a "superluminal group velocity" [33,88], our present analysis shows that it should rather be considered as a pure momentum property, while the group velocity is determined by the Poynting-Abraham energy flux and is always subluminal. The frequency dependence of the ratio of the SPP momentum and energy, Eq. (3.9), is shown in Fig. 3(b).

According to Eq. (2.11), the *kinetic* form of the Minkowski-type momentum, Eq. (2.8), has the same integral value (3.9), but its local density does not exhibit the nice proportionality to the energy density as in Eq. (3.9):

$$\tilde{\mathcal{P}}_M = g|A|^2 \frac{k_0^2}{k_p} \bar{\mathbf{z}} \begin{cases} \exp(-2\kappa_1 x), & x > 0 \\ \dfrac{1-\varepsilon+2\varepsilon^2}{\varepsilon(1+\varepsilon)} \exp(2\kappa_2 x), & x < 0 \end{cases} \qquad (3.10)$$

Moreover, this momentum density coincides with the Poynting-Abraham one in the vacuum. Therefore, for $x > 0$, $c\tilde{\mathcal{P}}_M / \tilde{W} = c\mathcal{P}_A / \tilde{W} = k_0 / k_p < 1$, and the kinetic Minkowski-type momentum *cannot explain the local super-momentum density in the vacuum evanescent field*, $c\tilde{\mathbf{P}}_M / \tilde{W} = k_p / k_0 > 1$, which is described by the canonical momentum (3.9) and is observed experimentally [31–33]. Comparing Eqs. (3.8)–(3.10), together with the singular character of the canonical Abraham-type momentum $\mathbf{P}_A$ [65], confirms that the *canonical* picture is more natural for the description of the *Minkowski*-type wave momentum, while the *kinetic* approach is more suitable for the characterization of the *Abraham*-type energy fluxes, see Table I.

*3.3. Spin and orbital AM of SPPs.*

We are now in the position to determine the spin and orbital AM of SPPs. Akin to the momentum of SPPs, these should be described using the Minkowski-type canonical picture. However, for completeness and comparison with other approaches, we first calculate the Abraham-type spin density (2.6) and its integral value. With the SPP fields (3.2) and (3.3), we obtain

$$\mathbf{S}_A = g|A|^2 \overline{\mathbf{y}} \begin{cases} \dfrac{\kappa_1}{k_p} \exp(-2\kappa_1 x), & x > 0 \\ -\dfrac{\kappa_2}{\varepsilon^2 k_p} \exp(2\kappa_2 x), & x < 0 \end{cases}, \qquad (3.11)$$

$$\langle \mathbf{S}_A \rangle = \frac{(-1-\varepsilon)\sqrt{-\varepsilon}}{(1+\varepsilon^2)} \frac{\langle \tilde{W} \rangle}{\omega} \overline{\mathbf{y}}. \qquad (3.12)$$

This is the *transverse helicity-independent spin*, first described in [65] and now attracting considerable attention [35,41–43,46,66–68]. We wrote Eq. (3.11) using the $\kappa_{1,2}/k_p$ factors to conform with the known results for the transverse spin in an evanescent wave in free space [35,41–43]: $S_{Ay} = \dfrac{\kappa_1}{k_p} \dfrac{\tilde{W}}{\omega}$ for $x > 0$. Equation (3.11) shows that the Abraham-type spin density $\mathbf{S}_A$ is *discontinuous* at the interface $x = 0$. As a result of this, the canonical and spin parts of the Abraham-Poynting energy flux have delta-function singularities [65], originating from the gradient terms in Eq. (2.5). Note also that the integral value $\langle S_{Ay} \rangle$ is always positive and is in agreement with calculations of Ref. [65] up to a factor of 2 missing there. [The missing factor of 2 in [65] originates from the improper application of the relation $\langle \mathbf{S}_A \rangle = \int (\mathbf{r} \times \mathbf{P}_A^s) dx = \dfrac{1}{2} \int \mathbf{r} \times (\nabla \times \mathbf{S}_A) dx$ to the $z$-delocalized SPP wave, involving only the term $\partial S_{Ay}/\partial x$ under the integral. In fact, this relation is valid only for localized wave packets involving two terms: $\partial S_{Ay}/\partial x$ and $\partial S_{Ay}/\partial z$.] At the same time, calculations of the integral Abraham spin "per particle" for surface Maxwell modes in [89] are not applicable in the SPP case because a dispersion-free model without proper Brillouin energy (2.3) was considered there. The frequency dependence of the Abraham-type spin (3.12) (in units of $\hbar$ "per plasmon") is shown in Fig. 4(a).

We now calculate the properly defined canonical Minkowski-type spin and orbital AM (2.13). Substituting Eqs. (3.2) and (3.3) into Eq. (2.13), we find

$$\tilde{\mathbf{S}}_M = g|A|^2 \overline{\mathbf{y}} \begin{cases} \dfrac{\kappa_1}{k_p} \exp(-2\kappa_1 x), & x > 0 \\ -\dfrac{\kappa_2}{k_p} \dfrac{2-\varepsilon}{\varepsilon^2} \exp(2\kappa_2 x), & x < 0 \end{cases} \qquad (3.13)$$

The spin AM density (3.13) is directed oppositely in the vacuum and metal: $S_y < 0$ for $x < 0$. This agrees with the opposite direction of rotation of the electric field **E** in the metal, Eq. (3.2a), but is in contrast to what is obtained for the "naïve" Minkowski spin (2.10), positive in the metal: $S_{My} > 0$ for $x < 0$ [72–74]. [Here we do not show the distribution of $\mathbf{S}_M$, Eq. (2.10), and only note that it is *continuous* at the interface $x = 0$; this assures the non-singular character of the canonical momentum $\tilde{\mathbf{P}}_M$, Eq. (2.9), which does not have any gradient corrections.] This proves that taking into account the dispersion-related corrections is crucial for determining the transverse spin and other dynamical properties of light in a dispersive medium. The integral expectation value of the spin AM (3.13) becomes:

$$\boxed{\langle \tilde{\mathbf{S}}_M \rangle = \frac{(-2-\varepsilon)\sqrt{-\varepsilon}}{1+\varepsilon^2} \frac{\langle \tilde{W} \rangle}{\omega} \bar{\mathbf{y}}} \ . \tag{3.14}$$

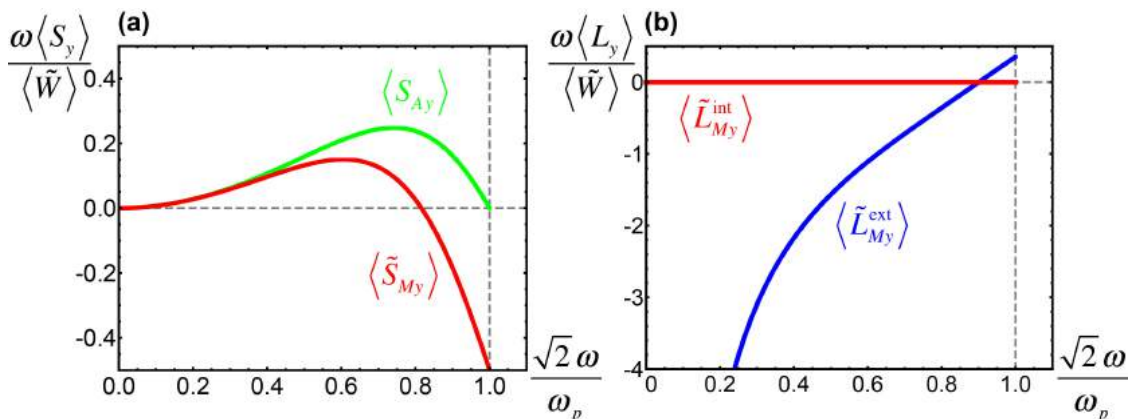

**Fig. 4**. (a) The integral Abraham-type transverse spin (3.12) and proper canonical Minkowski-type transverse spin (3.14) of the SPP versus frequency. In contrast to the positive Abraham-type spin, the Minkowski-type spin can have different directions and vanishes for $\omega = \omega_p/\sqrt{3}$. (b) The frequency dependences of the intrinsic (3.15) and extrinsic (with respect to the interface $x = 0$) (3.16) orbital angular momentum (AM) of the SPP. The vanishing intrinsic part means that the canonical momentum density (3.9) does not exhibit any vortex-like circulation, in contrast to the Poynting-Abraham energy flux (3.7) [24,65,85]. In turn, the extrinsic orbital AM originates from the shifted energy centroid of the SPP, Eq. (3.6) and Fig. 2(b).

Equation (3.14) provides *the first accurate calculation of the transverse spin AM carried by a SPP*. In contrast to the Abraham-like spin (3.12) considered before, its direction can vary depending on the SPP frequency. Namely, it is positive, $\langle \tilde{S}_{My} \rangle > 0$, for $\omega < \omega_p/\sqrt{3}$ and negative, $\langle \tilde{S}_{My} \rangle < 0$, for $\omega_p/\sqrt{3} < \omega < \omega_p/\sqrt{2}$, so that it *vanishes* at $\omega = \omega_p/\sqrt{3}$. The frequency dependence of the Minkowski-type spin (3.14) (in units of $\hbar$ per plasmon) is shown in Fig. 4(a). Note that its absolute value never exceeds $\hbar/2$ per plasmon; this is because of the pure-*electric* origin of the transverse spin, with no magnetic part. Interestingly, the critical zero-spin value $\omega = \omega_p/\sqrt{3}$, or $\varepsilon = -2$, corresponds to the elliptical $(x,z)$-polarizations of the electric field (3.2) with the axes ratios $\sqrt{2}$ and $1/\sqrt{2}$ (i.e., identical ellipticities but different orientations), and opposite directions of the rotation, in the vacuum and metal, respectively.

The orbital AM density in the SPP field is determined by Eq. (2.13) and the canonical momentum density (3.9): $\tilde{\mathbf{L}}_M = \mathbf{r} \times \tilde{\mathbf{P}}_M$. This quantity is *extrinsic*, i.e., depends on the choice of the coordinate origin. However, the expectation value of the orbital AM can have both *extrinsic* and *intrinsic* contributions [41]. The intrinsic orbital AM is calculated with respect to the centroid of the energy density distribution. Using the $x$-shifted centroid (3.6) and $z$-directed momentum density (3.9), we calculate the $y$-directed intrinsic orbital AM of the SPP:

$$\boxed{\left\langle \tilde{\mathbf{L}}_M^{\text{int}} \right\rangle = -\bar{\mathbf{y}} \int (x - \langle x \rangle) \tilde{P}_{Mz} \, dx = 0} \; . \tag{3.15}$$

Thus, the intrinsic orbital AM of SPP *vanishes*. This is because of the proportionality between the canonical momentum density (3.9) and energy density (3.4), which in turn determines the centroid (3.6). The vanishing intrinsic orbital AM (3.15) reflects the *non-vortex* character of the canonical momentum (3.9), in contrast to the circulating Abraham-type energy fluxes [24,65,85]. The extrinsic part of the orbital AM, calculated with respect to the interface $x = 0$, can be written as

$$\left\langle \tilde{\mathbf{L}}_M^{\text{ext}} \right\rangle = \left\langle \tilde{\mathbf{L}}_M \right\rangle - \left\langle \tilde{\mathbf{L}}_M^{\text{int}} \right\rangle = -\bar{\mathbf{y}} \langle x \rangle \langle \tilde{P}_{Mz} \rangle = -\frac{(1 + \varepsilon^2 + \varepsilon^3)}{2(1 + \varepsilon^2)\sqrt{-\varepsilon}} \frac{\langle \tilde{W} \rangle}{\omega} \bar{\mathbf{y}} \; . \tag{3.16}$$

This quantity can change its sign depending on the sign of $\langle x \rangle$. The frequency dependences of the intrinsic and extrinsic parts of the Minkowski-type orbital AM, Eqs. (3.15) and (3.16), (in units of $\hbar$ per plasmon) are shown in Fig. 4(b).

Below we examine the microscopic model of fields and electrons in the metal, which confirms the above phenomenological calculations and Minkowski-type picture of the momentum and AM of SPPs.

## 4. Microscopic calculations for a surface plasmon-polariton

*4.1. Microscopic fields and parameters of electron plasma.*

The microscopic approach is based on the separation of the microscopic electromagnetic field ($\mathbf{E}, \mathbf{H}$) and charges/currents inside the medium. In our case, the metal can be described using the Bloch hydrodynamic model for electron plasma. In this model, the electron density is written as $n(\mathbf{r},t) = n_0 + \tilde{n}(\mathbf{r},t)$, where $n_0$ is the uniform unperturbed density of free electrons (which neutralizes positive charges of background motionless ions), and $\tilde{n}$ is a small perturbation of the electron density caused by the interaction with electromagnetic wave fields. The local electron velocity is given by $\mathbf{v}(\mathbf{r},t)$. Considering a monochromatic linear problem, we introduce complex amplitudes for perturbations of electron properties, $\tilde{n}(\mathbf{r},t) = \text{Re}\left[\tilde{n}(\mathbf{r}) e^{-i\omega t}\right]$ and $\mathbf{v}(\mathbf{r},t) = \text{Re}\left[\tilde{\mathbf{v}}(\mathbf{r}) e^{-i\omega t}\right]$, and the time derivatives become $\partial/\partial t \to -i\omega$, entirely similar to complex field amplitudes. Microscopic electromagnetic fields always occur in free space (i.e., there is no effective medium, $\varepsilon = \mu = 1$), but the Maxwell equations are modified by the presence of charges and currents [22]:

$$\nabla \cdot \mathbf{H} = 0, \quad \mathbf{H} = -\frac{i}{k_0} \nabla \times \mathbf{E} \; ,$$

$$\nabla \cdot \mathbf{E} = 4\pi e \tilde{n}, \quad \mathbf{E} = \frac{i}{k_0} \nabla \times \mathbf{H} - i \frac{4\pi e n_0}{\omega} \tilde{\mathbf{v}} \; . \tag{4.1}$$

These equations describe the influence of the electrons on the fields. The back action is described by the hydrodynamic equation for the electron gas [71]:

$$-i\omega n_0 m \tilde{\mathbf{v}} = e n_0 \mathbf{E} - m\beta^2 \nabla \tilde{n}. \tag{4.2}$$

This equation is a classical equation of motion of the electron in the electric field $\mathbf{E}$ (the Lorentz force from the magnetic field vanishes in the linear problem) with the additional quantum pressure term involving the coefficient $\beta^2 = (3/5)v_F^2$, where $v_F$ is the Fermi velocity of electrons. Our classical treatment of the SPP wave implies the limit $\beta^2 \to 0$. However, we cannot omit the last term in Eq. (4.2) from the beginning because it is crucial to satisfy the boundary conditions at the metal-vacuum discontinuity, $x = 0$.

Solving Eqs. (4.1) and (4.2) with standard boundary conditions at the metal-vacuum interface (continuity of $E_x$, $E_z$, $H_y$, and vanishing of $v_x$) yields the microscopic electric and magnetic fields as well as electron plasma properties in the SPP wave. The magnetic field (both in the vacuum and in the metal) and electric field in the vacuum are still described by the macroscopic Eqs. (3.2) with parameters (3.1) and (3.3), while the electric field inside the metal becomes:

$$\mathbf{E} = \frac{A}{\varepsilon}\left\{\left[-(1-\varepsilon)e^{\gamma x} + e^{\kappa_2 x}\right]\bar{\mathbf{x}} + i\left[-(1-\varepsilon)\frac{k_p}{\gamma}e^{\gamma x} + \frac{\kappa_2}{k_p}e^{\kappa_2 x}\right]\bar{\mathbf{z}}\right\}\exp(ik_p z), \quad x < 0 \tag{4.3}$$

Here $\gamma^2 = k_p^2 - \varepsilon\omega^2/\beta^2$, we still use $\varepsilon$ as a parameter given by Eq. (3.1), and $\beta^2 \to 0$ implies $\gamma \to \infty$. The electron density and velocity perturbations in SPPs are given by

$$\tilde{n} = \frac{A}{4\pi e}\frac{\varepsilon-1}{\varepsilon}\left(\gamma - \frac{k_p^2}{\gamma}\right)e^{\gamma x}\exp(ik_p z), \quad x < 0, \tag{4.4}$$

$$\tilde{\mathbf{v}} = i\frac{A}{\varepsilon}\frac{e}{m\omega}\left[\left(-e^{\gamma x} + e^{\kappa_2 x}\right)\bar{\mathbf{x}} + i\left(-\frac{k_p}{\gamma}e^{\gamma x} + \frac{\kappa_2}{k_p}e^{\kappa_2 x}\right)\bar{\mathbf{z}}\right]\exp(ik_p z), \quad x < 0. \tag{4.5}$$

We can now consider the classical limit $\gamma \to \infty$. Since the $\gamma$-terms appear only in the metal half-space $x < 0$, one can use the limiting transition $\gamma\exp(\gamma x) \to \delta(x)$, where the delta-function describes *surface* effects at the interface. After doing so, the electric field (4.3) *becomes equal to the macroscopic one*, Eq. (3.2), while the electron density *vanishes* in the volume, $\tilde{n} = 0$ for $x < 0$, and acquires a surface delta-function singularity:

$$\tilde{n} = \frac{A}{4\pi e}\frac{\varepsilon-1}{\varepsilon}\exp(ik_p z)\delta(x). \tag{4.6}$$

As we show below, this singularity is cancelled by another singularity in field gradients, and all dynamical properties of the SPP wave are determined by *volume* contributions in the metal and in the vacuum. Finally, the electron velocity (4.5) becomes proportional to the electric field (3.2) [which follows from Eq. (4.2) at $\beta = 0$]:

$$\tilde{\mathbf{v}} = i\frac{A}{\varepsilon}\frac{e}{m\omega}\left(\bar{\mathbf{x}} + i\frac{\kappa_2}{k_p}\bar{\mathbf{z}}\right)\exp(ik_p z + \kappa_2 x) = \frac{ie}{m\omega}\mathbf{E}, \quad x < 0. \tag{4.7}$$

Note that it is the vanishing of the electron density perturbation $\tilde{n}$ in volume that makes the microscopic fields "transverse", i.e., divergence-free: $\nabla \cdot \mathbf{E} = 0$. Because of this, we do not need to consider contributions of "longitudinal" (i.e., curl-free) fields to the energy, momentum, and AM [90].

In addition to the point-charge features of electrons, we will need their electric-*dipole* properties. Since velocity is a time derivative of the position of the electron, we can write the complex amplitude of the electron displacement as $\tilde{\mathbf{a}} = \dfrac{i}{\omega}\tilde{\mathbf{v}}$. From here, the complex amplitude of the density of the electron dipole moment is

$$\tilde{\mathbf{d}} = n_0 e \tilde{\mathbf{a}} = \frac{i n_0 e}{\omega}\tilde{\mathbf{v}}. \tag{4.8}$$

Substituting here Eq. (4.7), we can write the dipole-moment density (4.8) as

$$\tilde{\mathbf{d}} = \alpha \mathbf{E}, \quad \alpha = -\frac{n_0 e^2}{m\omega^2} = \frac{\varepsilon - 1}{4\pi}. \tag{4.9}$$

Here $\alpha$ is the dipole polarizability of the electron gas, and the last equality shows that it is in perfect agreement with the macroscopic theory based on the permittivity $\varepsilon$ [22]. Indeed, substituting the velocity (4.7) into Maxwell equations (4.1) for microscopic fields, we immediately obtain the source-free Maxwell equations (2.2) for macroscopic fields with permittivity $\varepsilon$.

*4.2. Microscopic calculations of energy and momentum densities.*

In the vacuum half-space $x > 0$, the microscopic and macroscopic electromagnetic fields and their properties coincide, so we have to compare only the macroscopic and microscopic properties in the metal. Hereafter, we consider all quantities only in the $x < 0$ half-space.

The cycle-averaged energy density in the system of microscopic electromagnetic fields and electrons can be written as

$$\tilde{W} = \frac{g\omega}{2}\left(|\mathbf{E}|^2 + |\mathbf{H}|^2\right) + \frac{n_0 m |\tilde{\mathbf{v}}|^2}{4} \equiv W_0 + W_{\text{mat}}. \tag{4.10}$$

Here the first term is the microscopic-field energy [written as for free-space fields, Eq. (2.1)], and the second term is the kinetic energy of electrons. Note that the latter can also be presented as the energy of the dipole (4.8) and (4.9) in the electric field: $W_{\text{mat}} = -\dfrac{1}{4}\operatorname{Re}\left(\tilde{\mathbf{d}}^* \cdot \mathbf{E}\right) = -\dfrac{1}{4}\alpha|\mathbf{E}|^2$. Substituting fields (3.2) and velocity (4.7) into Eq. (4.10) results in the macroscopic energy density $\tilde{W}$, Eq. (3.4), at $x < 0$. Thus, the microscopic and macroscopic calculations are in perfect agreement. The field and electron contributions in the metal are:

$$W_0 = \frac{1+\varepsilon^2}{2(1-\varepsilon+\varepsilon^2)}\tilde{W}, \quad W_{\text{mat}} = \frac{(1-\varepsilon)^2}{2(1-\varepsilon+\varepsilon^2)}\tilde{W}. \tag{4.11}$$

The momentum density in the microscopic approach is also the sum of the field and electron contributions. The velocity of electrons in external fields is associated with the *kinetic* rather than canonical momentum of electrons, and, therefore, we consider the corresponding *kinetic* momentum density of the field. [It is worth noticing that the canonical electron momentum in the metal vanishes: $\mathbf{p} = m\tilde{\mathbf{v}} + \dfrac{e}{c}\mathbf{A} = m\tilde{\mathbf{v}} - i\dfrac{e}{\omega}\mathbf{E} = 0$, where we used the relation $\mathbf{A} = -i\dfrac{c}{\omega}\mathbf{E}$ for the transverse vector potential and Eq. (4.7).] For the microscopic field, the kinetic field momentum is given by the Poynting vector (1.1), i.e., the Abraham momentum. Below we show that adding it to the kinetic electron momentum yields the kinetic Minkowski-type momentum density (2.8) suggested by Philbin [16,17]:

$$\tilde{\mathcal{P}}_M = g k_0 \operatorname{Re}\left(\mathbf{E}^* \times \mathbf{H}\right) + \mathcal{P}_{\text{mat}} = \mathcal{P}_0 + \mathcal{P}_{\text{mat}}. \tag{4.12}$$

The calculation of the electron contribution $\mathcal{P}_{mat}$ requires a more sophisticated approach. Indeed, the simple expression $n_0 m \tilde{\mathbf{v}} = i \frac{e n_0}{\omega} \mathbf{E}$ (with oscillating, zero-average $\tilde{\mathbf{v}}$) for point electrons does not provide a meaningful result; instead of this, one has to consider an optical *force* acting on electric dipoles (4.8) and (4.9). We follow the formalism described in the review [4], Section 5.1 therein.

Namely, we consider a long but finite wave packet instead of a monochromatic continuous wave. Afterwards, the length of the wave packet can be tend to infinity. Introducing slowly-varying amplitudes $\mathbf{E}(\mathbf{r}) \to \mathbf{E}(\mathbf{r},t)$, $\mathbf{H}(\mathbf{r}) \to \mathbf{H}(\mathbf{r},t)$, etc., with the typical scale of the $t$-variations much larger than $\omega^{-1}$, involves the corresponding narrow but finite frequency Fourier spectrum centered at $\omega$. This produces the first-order Taylor-series correction to the relation (4.9) between the dipole moment and electric field:

$$\tilde{\mathbf{d}}(\mathbf{r},t) = \alpha(\omega) \mathbf{E}(\mathbf{r},t) + i \frac{d\alpha}{d\omega} \frac{\partial \mathbf{E}(\mathbf{r},t)}{\partial t}. \qquad (4.13)$$

Next, we consider the cycle-averaged force density acting on the dipole moment (4.13) in an external electromagnetic field [4]:

$$\mathbf{F} = \frac{1}{2} \operatorname{Re} \left[ \left( \tilde{\mathbf{d}}^* \cdot \nabla \right) \mathbf{E} + \tilde{\mathbf{d}}^* \times \left( \nabla \times \mathbf{E} \right) + \frac{1}{c} \frac{\partial}{\partial t} \left( \tilde{\mathbf{d}}^* \times \mathbf{H} \right) \right]. \qquad (4.14)$$

Substituting here Eq. (4.13), after some transformations the force can be written as:

$$\mathbf{F} = -\nabla W_{mat} + \frac{\partial}{\partial t} \left\{ \frac{1}{4} \frac{d\alpha}{d\omega} \operatorname{Im} \left[ \mathbf{E}^* \cdot (\nabla) \mathbf{E} \right] + \frac{1}{2c} \operatorname{Re} \left( \tilde{\mathbf{d}}^* \times \mathbf{H} \right) \right\}, \qquad (4.15)$$

where $W_{mat} = -\alpha |\mathbf{E}|^2 / 4$. The first term in Eq. (4.15) represents the gradient force, while the two terms subject to the time derivative should be associated with the momentum density carried by the electrons, i.e., $\mathcal{P}_{mat}$ of Eq. (4.12). Expressing the dipole-moment density and polarizability via the electric field and $\varepsilon$, Eq. (4.9), we arrive at:

$$\mathcal{P}_{mat} = \frac{g\omega}{2} \frac{d\varepsilon}{d\omega} \operatorname{Im} \left[ \mathbf{E}^* \cdot (\nabla) \mathbf{E} \right] + (\varepsilon - 1) g k_0 \operatorname{Re} \left( \mathbf{E}^* \times \mathbf{H} \right). \qquad (4.16)$$

Here the second term is associated with the "Abraham force" [22].

Substituting the electron momentum density (4.16) into Eq. (4.12), results in the kinetic form of the Minkowski-type momentum density (2.8) (for $\mu = 1$):

$$\tilde{\mathcal{P}}_M = \underbrace{\varepsilon g k_0 \operatorname{Re} \left( \mathbf{E}^* \times \mathbf{H} \right)}_{\mathcal{P}_0} + \frac{g\omega}{2} \frac{d\varepsilon}{d\omega} \operatorname{Im} \left[ \mathbf{E}^* \cdot (\nabla) \mathbf{E} \right]. \qquad (4.17)$$

Thus, using microscopic calculations for the SPP, we rigorously *derived* the kinetic momentum density (2.10) suggested previously from a phenomenological formalism [16,17].

We now trace the decomposition of the kinetic momentum (4.17) into the canonical (orbital) and spin parts (2.9) and (2.10). In principle, one substitutes the electron velocity $\tilde{\mathbf{v}} = (ie/m\omega) \mathbf{E}$ into the microscopic Maxwell equations (4.1), which results in the macroscopic Maxwell equations (2.2) with permittivity $\varepsilon$. Then, the decomposition becomes straightforward as described in Section 2. However, it is instructive to trace this decomposition at the microscopic level. For this purpose, we decompose the Poynting-like part of the kinetic momentum (4.17) into canonical and orbital parts using the microscopic Maxwell equations (4.1). The presence of sources, $\tilde{n}$ and $\tilde{\mathbf{v}}$, modifies this decomposition as compared to the free-space case (1.6)–(1.8):

$$\mathcal{P}_0 = \underbrace{\mathbf{P}_0 - 2\pi g \frac{n_0 e}{c}\operatorname{Im}(\mathbf{H}^* \times \tilde{\mathbf{v}})}_{\text{canonical}} + \underbrace{\frac{1}{2}\nabla \times \mathbf{S}_0 - 2\pi g e \operatorname{Im}(\mathbf{E}^*\tilde{n})}_{\text{spin}}. \tag{4.18}$$

Here we ascribed the two source-related terms to the canonical and spin parts such that the final result will coincide with the macroscopic equations.

First, for the SPP fields (3.2) and velocity (4.7) the velocity-related term contains only the canonical-type contribution [because $\operatorname{Im}(\mathbf{H}^* \times \mathbf{H}) = 0$ in SPPs]:

$$-\frac{n_0 e}{c}\operatorname{Im}(\mathbf{H}^* \times \tilde{\mathbf{v}}) = \frac{1-\varepsilon}{4\pi\varepsilon}\operatorname{Im}\left[\mathbf{H}^* \cdot (\nabla)\mathbf{H}\right]. \tag{4.19}$$

Combining the canonical part of Eq. (4.18) with Eq. (4.19) and the second term in Eq. (4.17), we obtain the *macroscopic canonical Minkowski-type momentum* (2.9) (with $\mu = 1$):

$$\tilde{\mathbf{P}}_M = \frac{g}{2}\operatorname{Im}\left[\tilde{\varepsilon}\mathbf{E}^* \cdot (\nabla)\mathbf{E} + \mathbf{H}^* \cdot (\nabla)\mathbf{H}\right], \tag{4.20}$$

Second, the curl of the free-space-like spin $\mathbf{S}_0$ for the SPP fields (3.2) contains a delta-function singularity at the metal-vacuum interface $x = 0$. Remarkably, this singularity is exactly cancelled by the singularity (4.6) in the electron density distribution, so that the spin part in Eq. (4.18) becomes *non-singular*. This confirms the non-singular character of the spin-orbital decomposition in the Minkowski-type approach, in contrast to the Abraham one [65]. In the bulk, $x < 0$, the spin part of Eq. (4.18) substituted in Eq. (4.17) immediately yields the corresponding part of the macroscopic Minkowski-type momentum involving the "naïve" Minkowski spin, Eq. (2.10):

$$\mathbf{P}_M^S = \frac{1}{2}\nabla \times \mathbf{S}_M, \quad \mathbf{S}_M = \frac{g}{2}\operatorname{Im}(\varepsilon \mathbf{E}^* \times \mathbf{E}). \tag{4.21}$$

Thus, we have obtained the macroscopic Minkowski-type momentum densities (2.8)–(2.10), both kinetic and canonical, using microscopic calculations in the metal with separated field and matter contributions.

*4.3. Microscopic approach to the spin and orbital AM.*

It might seem that the description of the AM quantities in dispersive media, given in Section 2, is somewhat "inconsistent" with the corresponding momentum quantities. Indeed, the Minkowski-type kinetic AM (2.12) is not simply determined by the corresponding kinetic momentum, $\mathbf{r} \times \tilde{\mathcal{P}}_M$, but contains additional dispersion-related terms. Furthermore, the spin momentum (2.10) is determined by the naïve Minkowski spin density $\mathbf{S}_M$, while the proper canonical spin AM $\tilde{\mathbf{S}}_M$, Eq. (2.13), differs from it in a dispersive medium. Importantly, the microscopic approach sheds light on these "inconsistencies".

Namely, the local motion of electrons provides an intrinsic contribution to the AM density [71], in fact, to its spin part. Using the electron displacement $\tilde{\mathbf{a}}$, Eq. (4.8), and velocity $\tilde{\mathbf{v}}$, Eq. (4.7), one can write this part of the AM density as:

$$\mathbf{S}_{\text{mat}} = \frac{n_0 m}{2}\operatorname{Re}(\tilde{\mathbf{a}}^* \times \tilde{\mathbf{v}}) = \frac{n_0 e^2}{2m\omega^3}\operatorname{Im}(\mathbf{E}^* \times \mathbf{E}) = \frac{g\omega}{2}\frac{d\varepsilon}{d\omega}\operatorname{Im}(\mathbf{E}^* \times \mathbf{E}). \tag{4.22}$$

This term exactly describes the dispersion-related addition in the Minowski-type kinetic AM (2.12) for the SPP wave:

$$\tilde{\mathcal{J}}_M = \mathbf{r} \times \tilde{\mathcal{P}}_M + \mathbf{S}_{\text{mat}}. \tag{4.23}$$

Thus, akin to the momentum density (2.8) and (4.175), microscopic calculations justify the kinetic AM density (2.12), previously obtained by Philbin within a phenomenological approach [17].

Consider now the spin and orbital AM densities. The orbital AM density is straightforwardly determined by the canonical momentum (4.20): $\tilde{\mathbf{L}}_M = \mathbf{r} \times \tilde{\mathbf{P}}_M$, and we have already described its properties in Section 3. At the same time, the intrinsic electron contribution (4.22) elucidate the difference between the naïve Minkowski spin density (4.21) and canonical spin density (2.13):

$$\tilde{\mathbf{S}}_M = \mathbf{S}_M + \mathbf{S}_{\text{mat}} = \frac{g}{2}\text{Im}\left(\tilde{\varepsilon}\mathbf{E}^* \times \mathbf{E}\right). \quad (4.24)$$

This justifies the use of the canonical spin AM in dispersive media and the transverse spin of a SPP calculated in Eqs. (3.13) and (3.14), Fig. 4(a). The dispersion-related contribution is absolutely crucial in the case of SPPs, because $\varepsilon < 0$, $\tilde{\varepsilon} = 2 - \varepsilon > 0$, and it changes both the magnitude and *sign* of the spin AM density in the metal.

*4.4. Magnetization, magnetization current, and Abraham momentum.*

The electron contribution (4.22) to the spin AM of a SPP corresponds to the microscopic *circular motion* of electrons in the SPP field. This microscopic *orbital* motion of the electrons produces multiple circulating currents, and, hence, the constant (non-oscillating) *magnetization* of the metal. Using the standard gyromagnetic ratio, we obtain the magnetization density in the metal:

$$\mathbf{M} = \frac{e}{2mc}\mathbf{S}_{\text{mat}} = \frac{ge\omega}{4mc}\frac{d\varepsilon}{d\omega}\text{Im}\left(\mathbf{E}^* \times \mathbf{E}\right). \quad (4.25)$$

This equation exactly coincides with the results [91,92] obtained for the magnetization of plasma by electromagnetic radiation and the *inverse Faraday effect* [22]. For the SPP fields (3.2) and (3.3), we find

$$\mathbf{M} = g|A|^2 \frac{-e}{2mc} \frac{2(1-\varepsilon)\sqrt{-\varepsilon}}{\varepsilon^2} \exp(2\kappa_2 x)\overline{\mathbf{y}}. \quad (4.26)$$

Thus, the metal is magnetized along the positive-$y$ direction ($e < 0$).

The magnetization (4.26) means that the SPP, being a mixed photon-electron excitation, carries a non-zero *magnetic moment*. To characterize this magnetic moment "per plasmon", we calculate the integral magnetization (4.26):

$$\boxed{\langle \mathbf{M} \rangle = \frac{-e}{2mc} \frac{2\sqrt{-\varepsilon}}{1+\varepsilon^2} \frac{\langle \tilde{W} \rangle}{\omega} \overline{\mathbf{y}}}. \quad (4.27)$$

This corresponds to the magnetic moment $\boldsymbol{\mu} = \frac{2\sqrt{-\varepsilon}}{1+\varepsilon^2}\mu_B \overline{\mathbf{y}}$ per plasmon, where $\mu_B = |e|\hbar/2mc$ is the Bohr magneton. The absolute value of this magnetic moment grows from 0 to $\mu_B$ as the SPP frequency $\omega$ changes from 0 to $\omega_p/\sqrt{2}$.

Moreover, the inhomogeneous magnetization (4.26) generates the corresponding magnetization *electric current* $\mathbf{j}_{\text{magn}} = c\nabla \times \mathbf{M}$:

$$\mathbf{j}_{\text{magn}} = g|A|^2 \frac{-e}{m} \frac{2(1-\varepsilon)\sqrt{-\varepsilon}}{\varepsilon^2} \kappa_2 \exp(2\kappa_2 x)\overline{\mathbf{z}}. \quad (4.28)$$

This is a *direct current* which flows in the metal in the $z$ direction, i.e., *along* the SPP propagation. It should be emphasized that the current (4.28) is obtained as a *quadratic* form of the SPP fields. Indeed, the linear-approximation current is determined by the electron velocity $\tilde{\mathbf{v}}$ and vanishes after cycle averaging.

We also note that the magnetization current is solenoidal (divergenceless). Therefore, it does not contribute to the charge transport and cannot be measured by an ammeter or voltmeter. Nonetheless, one can determine the electron velocity $\mathbf{v}_{\text{magn}} = \mathbf{j}_{\text{magn}} / en_0$ and momentum density $\boldsymbol{\mathcal{P}}_{\text{magn}} = mn_0 \mathbf{v}_{\text{magn}} = (m/e) \mathbf{j}_{\text{magn}}$ corresponding to the direct current (4.28). Using Eqs. (3.3), we write it as

$$\boldsymbol{\mathcal{P}}_{\text{magn}} = -g|A|^2 \frac{k_0^2}{k_p} \frac{2(1-\varepsilon)}{-1-\varepsilon} \exp(2\kappa_2 x) \bar{\mathbf{z}}. \qquad (4.29)$$

Thus, the magnetization momentum (4.29) is directed *oppositely* to the SPP propagation. Most significantly, it is exactly equal to the difference between the kinetic Abraham and Minkowski-type momentum densities in the metal, Eqs. (3.7) and (3.10):

$$\boxed{\boldsymbol{\mathcal{P}}_A = \tilde{\boldsymbol{\mathcal{P}}}_M + \boldsymbol{\mathcal{P}}_{\text{magn}}}. \qquad (4.30)$$

Equation (4.30) completes the microscopic picture and explains the origin of the difference between the Abraham and Minkowski momenta in the medium. Since this difference is produced by the *direct* magnetization current, it cannot be attributed to the *wave* (Minkowski-type) momentum but it does contribute to the energy flux (Abraham-Poynting momentum) and the group velocity of SPPs. Note that the Abraham momentum density can also be obtained as a pure *microscopic-field momentum* $\boldsymbol{\mathcal{P}}_A = \boldsymbol{\mathcal{P}}_0$, see Eq. (4.12) [19]. It follows from here that the total momentum of the metal *vanishes* in the problem under consideration: the electron contribution to the wave momentum is exactly cancelled by the magnetization-current momentum: $\boldsymbol{\mathcal{P}}_{\text{mat}} + \boldsymbol{\mathcal{P}}_{\text{magn}} = 0$.

Interestingly, recent work [93] claimed that the magnetization of the medium should be twice as small as compared to Eq. (4.22) due to the contribution from a "drift current". If this were to be the case, then adding the corresponding magnetic current and momentum (4.29) to the Minkowski-type momentum would yield an average between the Minkowski and Abraham momenta in Eq. (4.30), similar to the suggestion by Mansuripur [6,94].

## 5. Helicity and duality aspects

*5.1. Helicity density and flux in a medium.*

Here we briefly consider problems related to the *dual symmetry* between electric and magnetic fields in Maxwell equations [52,54–57,69,70,75,76,79]. This symmetry is exact in free-space Maxwell equations, but the presence of *electric* charges and currents in matter breaks it. The dual symmetry corresponds to the conservation of electromagnetic *helicity* via Noether's theorem [55,57,69,70,75,76,95–98]. Using the same phenomenological Lagrangian formalism as for the derivation of Minkowski-type momentum and AM densities (2.8) and (2.12), Philbin obtained the *helicity density* in a dispersive medium [77]:

$$\tilde{K} = \frac{g}{2}\left(\frac{\tilde{\varepsilon}}{\varepsilon} + \frac{\tilde{\mu}}{\mu}\right) \text{Im}(\mathbf{E} \cdot \mathbf{H}^*). \qquad (5.1)$$

In a dispersion-free medium, $\tilde{\varepsilon} = \varepsilon$ and $\tilde{\mu} = \mu$, and Eq. (5.1) yields the free-space result $K = g \,\text{Im}(\mathbf{E} \cdot \mathbf{H}^*)$ [55,57,59,69]. It should be emphasized that $\tilde{K}$ is the helicity density, but not the chirality density (Lipkin's zilch); these quantities are simply proportional to each other only in free space [57,77,99–103]. Similarly to Eqs. (2.17), for a plane wave in a homogeneous transparent medium, Eq. (5.1) yields

$$\frac{\tilde{K}}{\tilde{W}} = \frac{1}{n_p}\frac{\sigma}{\omega}. \tag{5.2}$$

This means that the helicity "per photon" in units of $\hbar$ is limited in the medium by the $\left(-n_p^{-1}, n_p^{-1}\right)$ range. This is a strange result without clear physical meaning. Using the quantum wavefunction formalism (2.15) and (2.16) with the helicity operator [55,59] $\hat{K} = \begin{pmatrix} 0 & i \\ -i & 0 \end{pmatrix}$ would produce a more natural result

$$\frac{\tilde{K}}{\tilde{W}} = \frac{\sigma}{\omega}, \tag{5.3}$$

but only in the case of a non-dispersive medium. Therefore, the helicity density in a dispersive medium is still controversial and requires further investigation. In any case, for the SPP fields (3.2), the helicity vanishes [65] due to the orthogonality of the electric and magnetic fields.

Note also that in free space the dual-symmetric spin AM density (1.7) determines the *helicity flux* density [55,57,69,70,100]. In a dispersive medium, calculations in Ref. [77] showed that the flux density of the helicity (5.1) is determined by the *Abraham* spin density $\mathbf{S}_A$, Eq. (2.6). In this case, the helicity flux density does not coincide with the proper Minkowski-type spin AM density (2.13) in a medium, and these are two different physical quantities. Even considering the Lipkin's *chirality* density, its flux becomes proportional to the "naïve" Minkowski spin density $\mathbf{S}_M$ [77,102], which is different from the canonical spin density (2.13) in dispersive media.

*5.2. Dual-symmetric and asymmetric quantities.*

So far, we considered all definitions of the optical energy, momentum, and AM using forms symmetric with respect to the electric and magnetic fields (and, correspondingly, indices $\varepsilon$ and $\mu$) [35,41,42,46,52,54–57,59,65,69,70]. However, in standard electromagnetic field theory (or QED) the field Lagrangian is *not* dual-symmetric [36,44,55,69,70]. Due to this, the canonical momentum, spin, and orbital AM densities are often defined using dual-asymmetric field-theory expressions, which contain only the electric-field parts [36,44,47–50,60,90]. In such "standard" formalism in free space, the energy density and kinetic Poynting momentum density are still given by the dual-symmetric expressions (1.1) and (2.1), whereas the canonical momentum, spin, and spin AM densities (1.6)–(1.8) become [54,55]:

$$\mathbf{P}_0' = 2\mathbf{P}_0^e = g\,\mathrm{Im}\left[\mathbf{E}^* \cdot (\nabla)\mathbf{E}\right], \tag{5.4}$$

$$\mathbf{L}_0' = \mathbf{r} \times \mathbf{P}_0', \quad \mathbf{S}_0' = 2\mathbf{S}_0^e = g\,\mathrm{Im}\left[\mathbf{E}^* \times \mathbf{E}\right]. \tag{5.5}$$

Here, we introduced the electric and magnetic parts of the momentum and spin densities (1.6) and (1.7): $\mathbf{P}_0 = \mathbf{P}_0^e + \mathbf{P}_0^m$ and $\mathbf{S}_0 = \mathbf{S}_0^e + \mathbf{S}_0^m$. Adopting the definitions (5.4) and (5.5) would mean that the only the phase gradients of the *electric* (but not magnetic) field produce momentum and orbital AM, and only rotations of the *electric* (but not magnetic) field generate the spin AM. On the one hand, this is not satisfactory from a general physical perspective, where electromagnetic waves involve electric and magnetic fields on equal footing [52,55,57,56,81–83]. On the other hand, the electric-biased quantities (5.4) and (5.5) can be useful in some practical problems considering interactions of fields with *electric*-dipole particles or atoms, which are not sensitive to magnetic fields [47–49,59,60]. However, for magnetic-dipole particles, the electric-biased quantities (5.4) and (5.5) do not make sense.

Note that in free space, the dual-asymmetric densities (5.4) and (5.5) do not cause significant problems because: (i) the Poynting-vector decomposition still has the same form (1.8): $\mathcal{P}_0 = \mathbf{P}_0' + \frac{1}{2}\nabla \times \mathbf{S}_0'$, and (ii) the integral values of all these quantities for localized free-space fields coincide with the dual-symmetric definitions [52,55]:

$$\left\langle \mathbf{P}_0' \right\rangle = \left\langle \mathbf{P}_0 \right\rangle = \left\langle \mathcal{P}_0 \right\rangle, \quad \left\langle \mathbf{S}_0' \right\rangle = \left\langle \mathbf{S}_0 \right\rangle. \tag{5.6}$$

However, in a medium, which is dual-asymmetric ($\varepsilon \neq \mu$) in the generic case, the difference between the dual-symmetric and asymmetric definitions becomes crucial. We again consider the example of a SPP wave at a metal-vacuum interface. First, the electric and magnetic contributions to the energy density (2.3) and (3.4) for the macroscopic SPP fields (3.2) equal

$$\tilde{W}^e = \frac{g}{2}|A|^2 \omega \begin{cases} \dfrac{\varepsilon - 1}{\varepsilon}\exp(-2\kappa_1 x), & x > 0 \\ \dfrac{(\varepsilon - 1)(\varepsilon - 2)}{\varepsilon^2}\exp(2\kappa_2 x), & x < 0 \end{cases} \tag{5.7a}$$

$$\tilde{W}^m = \frac{g}{2}|A|^2 \omega \begin{cases} \dfrac{\varepsilon + 1}{\varepsilon}\exp(-2\kappa_1 x), & x > 0 \\ \dfrac{\varepsilon + 1}{\varepsilon}\exp(2\kappa_2 x), & x < 0 \end{cases} \tag{5.7b}$$

Obviously, these contributions are different and result in different integral electric and magnetic energy parts:

$$\left\langle \tilde{W}^e \right\rangle = \frac{2 - \varepsilon + \varepsilon^2}{2(1 + \varepsilon^2)}\left\langle \tilde{W} \right\rangle, \quad \left\langle \tilde{W}^m \right\rangle = \frac{\varepsilon(1 + \varepsilon)}{2(1 + \varepsilon^2)}\left\langle \tilde{W} \right\rangle. \tag{5.8}$$

Then, the electric and magnetic parts of the canonical Minkowski-type momentum (2.9) and (3.9) become proportional to the corresponding energy parts:

$$\tilde{\mathbf{P}}_M^{e,m} = k_p \frac{\tilde{W}^{e,m}}{\omega}\overline{\mathbf{z}}, \quad \left\langle \tilde{\mathbf{P}}_M^{e,m} \right\rangle = k_p \frac{\left\langle \tilde{W}^{e,m} \right\rangle}{\omega}\overline{\mathbf{z}}. \tag{5.9}$$

From here and the difference in the integral electric and magnetic energies (5.8), it follows that using an electric-biased momentum density similar to Eq. (5.4), $\tilde{\mathbf{P}}_M' = 2\tilde{\mathbf{P}}_M^e = g\,\mathrm{Im}\left[\tilde{\varepsilon}\mathbf{E}^* \cdot (\nabla)\mathbf{E}\right]$, would yield

$$\left\langle \tilde{\mathbf{P}}_M' \right\rangle = k_p \frac{2\left\langle \tilde{W}^e \right\rangle}{\omega}\overline{\mathbf{z}} \neq k_p \frac{\left\langle \tilde{W} \right\rangle}{\omega}\overline{\mathbf{z}} = \left\langle \tilde{\mathbf{P}}_M \right\rangle. \tag{5.10}$$

Thus, the dual-asymmetric definitions in a medium do *not* satisfy the convenient free-space relations (5.6). In the case of Eq. (5.10), this breaks the natural proportionality (3.9) and the momentum of the SPP becomes *not* equal to $\hbar k_p$ per plasmon. Obviously, this is a physically unsatisfactory result. Therefore, we conclude that *dual-symmetric definitions of momentum and AM densities are crucial for structured waves in inhomogeneous optical media*.

Moreover, the validity of the dual-symmetric (rather than electric-biased) formalism follows from the *microscopic* calculations of Section 4. Indeed, the dispersion-related material correction in Eq. (4.17) has the form $\omega \frac{d\varepsilon}{d\omega} \mathbf{P}_0^e$, independently of the formalism. It is naturally combined with the corresponding electric term $\varepsilon \mathbf{P}_0^e$ of the *dual-symmetric* spin-orbital decomposition of the first term of (4.17), yielding the electric part of the canonical dual-symmetric Minkowski-type momentum (2.9): $\tilde{\mathbf{P}}_M^e = \tilde{\varepsilon} \mathbf{P}_0^e$, see Eqs. (4.18)–(4.20). However, using the *electric-biased* decomposition (5.4) and (5.5) doubles the non-dispersive term: $2\varepsilon \mathbf{P}_0^e$, and then it *cannot* be combined with the same dispersive term into a meaningful result proportional to $\tilde{\varepsilon}$. The same situation occurs in the microscopic derivation of the spin AM. The dispersive material term $\omega \frac{d\varepsilon}{d\omega} \mathbf{S}_0^e$ (4.22) is naturally combined with the electric part of the *dual-symmetric* spin density $\mathbf{S}_M^e = \varepsilon \mathbf{S}_0^e$, producing the electric part of the canonical dual-symmetric Minkowski-type spin (2.13): $\tilde{\mathbf{S}}_M^e = \tilde{\varepsilon} \mathbf{S}_0^e$, see Eq. (4.24). In turn, the electric-biased non-dispersive spin $2\varepsilon \mathbf{S}_0^e$ cannot be combined with the fixed dispersive term. Thus, *the dual-symmetric forms of the canonical momentum and AM densities are justified on the microscopic level by fixed dispersive material terms in the momentum and AM densities*.

The transverse spin AM of SPPs is a quantity which is extremely sensitive to the duality. Indeed, for the metal-vacuum interface considered here, the transverse spin has a purely *electric* nature: only the electric field rotates in solutions (3.2), while the transverse magnetic field yields no contribution to the spin AM [65]. Using an electric-biased definition of the spin AM, similar to Eq. (5.5), $\tilde{\mathbf{S}}_M{'} = 2\tilde{\mathbf{S}}_M^e = g \operatorname{Im}(\tilde{\varepsilon} \mathbf{E}^* \times \mathbf{E})$, we would obtain a transverse spin twice as large as Eqs. (3.13) and (3.14): $\langle \tilde{\mathbf{S}}_M{'} \rangle = 2 \langle \tilde{\mathbf{S}}_M \rangle$. However, one could also consider surface waves at an interface between the vacuum and a magnetic medium with $\varepsilon = 1$ and $\mu < -1$. In this case, the surface wave would be given by Eqs. (3.2) with swapped electric and magnetic fields, and the transverse spin would have a purely *magnetic* nature. Obviously, the electric-biased definition would yield no spin AM at all: $\langle \tilde{\mathbf{S}}_M{'} \rangle = 0$. This is a clear evidence that the equality of integral electric and magnetic spins for localized states in free space [52] does not hold true in media:

$$\langle \tilde{\mathbf{S}}_M{'} \rangle \neq \langle \tilde{\mathbf{S}}_M \rangle. \qquad (5.11)$$

Overall, we conclude that only the *dual-symmetric* definitions of the momentum, spin, and orbital AM properties of optical fields in dispersive media provide physically consistent and satisfactory results.

## 6. Concluding remarks

We have examined momentum and angular-momentum (AM) properties of monochromatic optical fields in dispersive and inhomogeneous media. The two major problems that lie at the heart of this study are: (i) the Abraham-Minkowski dilemma and (ii) the canonical spin-orbital decomposition of optical momentum and AM. We have shown that, in principle, one can formulate four momentum-AM pictures using Abraham-type and Minkowski-type quantities in kinetic (i.e., Poynting-like, without spin-orbital separation) and canonical (i.e., with spin-orbital separation) approaches. These four pictures are summarized in Table I. However, two of these sets of quantities are more physically meaningful.

First, the Abraham-Poynting kinetic momentum density (1.1) and (1.2), $\mathcal{P}_A = \mathcal{P}_0$, should be associated with the *energy flux* density rather than the momentum density. This quantity determines the energy transport and *group velocity* of a wave packet in the medium.

Second, the canonical Minkowski-type momentum density (2.9), $\tilde{\mathbf{P}}_M$, together with the corresponding spin and orbital AM densities (2.13), $\tilde{\mathbf{S}}_M$ and $\tilde{\mathbf{L}}_M = \mathbf{r} \times \tilde{\mathbf{P}}_M$, provide a physically-meaningful and self-consistent description of the momentum and AM of light in the medium. To the best of our knowledge, these quantities were derived for the first time in the present work, but these are consistent with several previously used approaches. On the one hand, the kinetic counterpart of this canonical picture, momentum (2.8), $\tilde{\mathcal{P}}_M$, and total AM (2.12), $\tilde{\mathcal{J}}_M$, exactly coincide with those obtained in the most general form by Philbin and Allanson [16,17]. On the other hand, our canonical characteristics (2.9) and (2.13) have a much more elegant form, exhibiting a pleasing similarity with the Brillouin energy density (2.3), $\tilde{W}$, in the medium. As a result, the energy, momentum, spin, and orbital AM densities in the medium can be written in a laconic unified form (2.15) using the corresponding quantum-mechanical operators and proper inner product modified by the $\tilde{\varepsilon}$ and $\tilde{\mu}$ indices of the medium. This coincides with the general approach to electromagnetic bi-linear forms developed by Silveirinha [81–83].

We applied the above general theory to a surface plasmon-polariton (SPP) wave at a metal-vacuum interface. This example provides a deep physical insight because it involves essentially *inhomogeneous* fields as well as an *inhomogeneous*, *non-transparent*, and *dispersive* medium. This is in sharp contrast to plane waves in homogeneous transparent media, which are considered in the majority of the Abraham-Minkowski studies. We have shown that in the non-trivial SPP field, the integral Abraham-Poynting momentum $\langle \mathcal{P}_A \rangle$ describes the group velocity $v_g = \partial \omega / \partial k_p < c$ of SPPs, Eq. (3.8), while the integral canonical momentum $\langle \tilde{\mathbf{P}}_M \rangle$ corresponds to the *super-momentum* $\hbar k_p > \hbar k_0$ per plasmon, Eq. (3.9). This is the first example of a wave, which carries an integral momentum larger than that of a photon in vacuum, and this originates from the inhomogeneous-evanescent character of the surface wave rather than from the medium refractive index ($k_p \to \infty$ at $\varepsilon \to -1$ in the metal).

We have also provided the first accurate calculation of the *transverse spin* $\langle \tilde{\mathbf{S}}_M \rangle$ of a SPP, Eq. (3.14). The result differs considerably from previous calculations using the Abraham-type definition of the spin [65]. In particular, the integral transverse spin AM of a SPP can vanish (at $\omega = \omega_p / \sqrt{3}$), change its sign and reach the value $-\hbar/2$ per plasmon, Fig. 4(a). In turn the intrinsic orbital AM (calculated with respect to the center of energy) of a SPP *vanishes*, Eq. (3.15). This agrees with the non-vortex character of the canonical momentum density (phase gradient) in the SPP field, and is in contrast to the Abraham-Poynting circulating energy flux [65].

Thus, the SPP example shows that the Abraham-type kinetic and Minkowski-type canonical properties provide intuitively clear and consistent description of complex optical fields in complex media. Importantly, we have also provided *microscopic* calculations of the momentum and AM densities for the SPP field, considering the microscopic electromagnetic fields and motion of free electrons in the metal. These calculations resulted precisely in the Minkowski-type momentum and AM densities, previously suggested from macroscopic phenomenological approaches [16,17]. This proves the validity of our approach and illuminates its physical origin. Importantly, the dispersion $\varepsilon(\omega)$ in the metal was absolutely crucial in the above calculations, affecting not only the magnitudes but also the signs of the dynamical quantities (because of $\varepsilon < 0$ and $\tilde{\varepsilon} > 0$).

Using the microscopic theory, we have also predicted a *transverse magnetization* of the metal (the inverse Faraday effect) corresponding to the transverse spin of a SPP. This means that a SPP wave carries not only the spin AM but also the *transverse magnetic moment*, up to a Bohr magneton per plasmon. Furthermore, an inhomogeneous magnetization produces the *direct magnetization current* flowing along the metal surface in the SPP propagation direction. Remarkably, the momentum density corresponding to this current is exactly equal to the difference between the Abraham and Minkowski-type momenta in the metal, thereby providing one more "resolution" of this longstanding problem [1–7].

Finally, we briefly discussed the optical helicity and duality problems in dispersive media. The analysis of the SPP example leads us to conclude that the dual-symmetric description of the canonical momentum, spin and orbital AM, with symmetric electric- and magnetic-field contributions, is crucially important for the physically meaningful and consistent picture of these properties in dispersive media. In particular, we found that *microscopic* calculations, including dispersion-related material terms, are consistent with the *dual-symmetric* (rather than electric-biased) formalism.

Thus, the present study provides a complete analysis and description of the momentum and AM of light in dispersive and inhomogeneos (but isotropic and lossless) media. We have considered SPPs only as the simplest example of the application of our theory, where other approaches fail. Taking into account both material and structured-light properties is crucial in a variety of nanooptical and photonic systems, including photonic crystals, metamaterials, and optomechanical systems. Our theory provides an efficient toolbox for the description of dynamical properties of light in such systems. One of the main tasks for future studies is to extend this analysis to other classes of materials, including dissipation or gain and anisotropy. In particular, it is not clear if one can separate the spin and orbital degrees of freedom in anisotropic media. Close correspondence of our approach to some of the results of Refs. [18,81–83,102] (dealing with quite general bi-anisotropic media), suggests that the analysis presented in this work can be extended to more complex cases.

Another important direction for future consideration is whether the fundamental wave characteristics introduced in this work can be observed in experiments. There were experiments measuring Minkowski-type momentum ($\propto \hbar \mathbf{k}$) for *plane waves* in dispersive media [13,104,105]. In addition, the canonical momentum and spin AM densities in structured optical fields in *free space* are directly observable via the optical force and torque on dipole particles or atoms [31–35,41,46,47,59–63,66]. Therefore, now it would be important to calculate and measure optical forces and torques on small particles in dispersive media (e.g., liquids or gases), and check if these are proportional to the canonical momentum and spin densities in dispersive media. Furthermore, it would be important to find an experimental setup where the *integral* super-momentum and transverse spin of SPPs, derived in this work, could be detected. So far, only local densities of these quantities were accessible via local free-space measurements using small particles or atoms [31–35,41,66,67,106–109].

*Note added*. A short (letter) version of this work, highlighting its main results without derivations, has recently been published in [110]. After the submission of this work, we became aware of the recent preprint [111], where material corrections to the electromagnetic spin density, similar to those obtained in our work, were found in the form of a torque in the spin continuity equation.

## Acknowledgements

We acknowledge helpful discussions with T. G. Philbin, F. J. Rodríguez-Fortuño, I. Fernandez-Corbaton, M. Silveirinha, M. Masuripur, E. Leader, Y. P. Bliokh, M. Durach, and N. Noginova. This work was supported by the RIKEN iTHES Project, MURI Center for Dynamic Magneto-Optics via the AFOSR Award No. FA9550-14-1-0040, the Japan Society for the Promotion of Science (KAKENHI), the IMPACT program of JST, CREST grant No. JPMJCR1676, the John Templeton Foundation, the RIKEN-AIST "Challenge Research" program, and the Australian Research Council.